\documentclass[letterpaper,journal,final]{IEEEtran}

\newcommand{\subparagraph}{}

\usepackage{graphicx,psfrag}
\usepackage{amsmath,amssymb,amsfonts,mathrsfs,mathtools}
\usepackage{color,epstopdf}
\usepackage{algorithmic}
\usepackage{enumerate} 
\usepackage{subfigure}
\usepackage{floatrow}
\usepackage{float}
\usepackage{stfloats}
\usepackage{flushend}
\usepackage{bbold}
\usepackage{dsfont}
\usepackage{mathrsfs}
\usepackage{relsize}
\usepackage{relsize}

\newtheorem{assumption}{Assumption}
\newtheorem{definition}{Definition}
\newtheorem{theorem}{Theorem}
\newtheorem{lem}{Lemma}
\newtheorem{corollary}{Corollary}
\newtheorem{proposition}{Proposition}

\newcommand{\R}{\mathbb{R}}

\newcommand{\bbm}{\begin{bmatrix}}
\newcommand{\ebm}{\end{bmatrix}}

\newcounter{mytempeqncnt}

\def\be{\begin{equation}}
\def\ee{\end{equation}}
\def\ba{\begin{array}}
\def\ea{\end{array}}
\def\eqa{\begin{eqnarray}}
\def\eqe{\end{eqnarray}}

\definecolor{darkgreen}{rgb}{0.0, 0.55, 0.0}
\definecolor{amaranth}{rgb}{0.9, 0.17, 0.31}
\usepackage[prependcaption,colorinlistoftodos]{todonotes}

\begin{document}

\title{Learning Controllers from Data via Approximate Nonlinearity Cancellation}

\author{C. De Persis, M. Rotulo, and P. Tesi 
\thanks{C. De Persis and M. Rotulo are with ENTEG and the J. C. Willems Center for
Systems and Control, 
University of Groningen, 9747 AG Groningen, The Netherlands.
Email: {\tt\small \{c.de.persis,m.rotulo\}@rug.nl}. \\
P. Tesi is with DINFO, University of Florence, 50139 Firenze, Italy 
E-mail: {\tt\small pietro.tesi@unifi.it}.}
}

\maketitle
\begin{abstract}
We introduce a method to deal with the data-driven control design of nonlinear systems. 
We derive conditions to design controllers via (approximate) nonlinearity cancellation. 
These conditions take the compact form of data-dependent semi-definite programs. 
The method returns controllers that can be certified to stabilize the system even when 
data are perturbed and disturbances affect the dynamics of the system during the execution 
of the control task, in which case an estimate of the robustly positively invariant set is provided. 
\end{abstract}

\section{Introduction}\label{sec:intro}

\IEEEPARstart{A}utomating the control design process is important to cope 
with complex dynamical plants whose dynamics is poorly known. Data-driven control is 
a notable example of such an automated synthesis.
Namely, data-driven control refers to the procedure of designing controllers for an unknown 
system starting solely from measurements collected from the plant and some priors about the 
plant itself (linear vs.~nonlinear parametrization, nature of the noise, etc.). In this paper we study 
the problem of designing controllers for nonlinear systems from data. 

{\it Related literature.} System identification followed by control design for the identified system is a 
classical way to indirectly perform data-driven control \cite{pillonetto2014kernel}. 
By {\em direct} data-driven control instead it is meant a procedure in which no intermediate step 
of identifying the system model is taken, 
earlier examples being the iterative feedback tuning (IFT) \cite{HGGL98}, 
and the virtual reference feedback tuning (VRFT) \cite{CLS02}. 
Recent times have seen a renewed interest in direct data-driven control, viewed as compact data-dependent 
conditions which, once verified, automatically return controllers without {\em explicitly} identifying the plant. 
One of the focus points in these data-driven control results is how to deal with perturbations and noise 
affecting the data and the resulting noise-induced uncertainty. Assuming a process noise with bounded 
$\ell_\infty$ norm, \cite{dai2018moments} defines a set of system's matrices pairs consistent with the data
and, using an extended Farkas' lemma, 
derives conditions under which stability of all systems in the set hold. 
These conditions can be checked using polynomial optimization techniques. 

The papers \cite{JC-JL-FD:18, coulson2021distributionally}
highlight the relevance of a result in \cite{willems2005}, 
about representing the behavior of a linear time-invariant system via a 
single input-output trajectory, and use this result 
to develop {\em data-enabled}, rather than model-based, predictive control, 
providing probabilistic guarantees on performance for systems subject to stochastic disturbances.

The result of \cite{willems2005} has also been used in \cite{de2019formulas} to obtain 
a data-dependent representation for linear systems based on which linear matrix 
inequalities only depending on data are introduced and used to provide solutions 
to problems such a state- and output-feedback stabilization as well as the linear 
quadratic regulator synthesis. The presence of deterministic noise with bounded 
energy affecting the data is dealt with a matrix elimination result to get rid of the 
resulting noise-induced uncertainty in the representation. 

If the samples of process noise are i.i.d. and Gaussian, then \cite{dean2019sample} 
provides a quantification in probability of the confidence region, which \cite{umenberger2019robustLCSS} 
exploits to give data-dependent conditions for minimizing the worst case cost of the LQ problem 
over all the system's matrices in the confidence region. The technical tool for this 
study is an extension of the S-lemma provided in \cite{luo2004multivariate}. 
A new matrix S-lemma is introduced in \cite{henk-ddctr-uncer} to provide non-conservative
conditions for designing controllers from data affected by disturbances satisfying quadratic bounds. 
Other results to deal with disturbances use a full-block S-procedure and linear fractional 
representations \cite{berberich2020combining_}, the classical S-procedure \cite{BisoffiSCL2021tradeoffs} 
and Petersen's lemma \cite{Bisoffi2021Petersen}. 
%

The majority of the available results consider linear systems.
Unsurprisingly, deriving solutions for \emph{nonlinear} systems is harder. 
Earlier representative results of data-driven control of nonlinear systems include 
the nonlinear extension of VRFT \cite{campi2006direct}, the design of controllers 
in the form of kernel functions tuned using data via set-membership identification 
techniques \cite{tanaskovic2017data}, and
the so-called model-free control \cite{fliess2013model, tabuada2017data}.

A way  to deal with nonlinear systems is to exploit some structure, 
when it is a priori known the class to which the system belongs.
Data-driven control of second-order Volterra systems is studied 
in \cite{schifferCDC20volterra} and data-dependent LMI-based stabilization of bilinear systems 
in \cite{bisoffi2020data}, the latter being motivated by Carleman bilinearization of 
general nonlinear systems. A point-to-point optimal control problem for bilinear 
systems is formulated in the recent work \cite{yuan2021data}. The data-driven control 
design for polynomial systems is the subject of \cite{dai2020semi,guoTAC2021poly}. 
While \cite{dai2020semi} uses Rantzer's dual Lyapunov's theory and moments based techniques, 
\cite{guoTAC2021poly} uses Lyapunov second method
and a particular parametrization of the Lyapunov function to obtain SOS programs
whose feasibility directly provide stabilizing controllers. 
See \cite{Bisoffi2021Petersen} for additional results on the data-driven 
control design of polynomial systems based on Petersen's lemma. 
When the system is not polynomial, the approach in \cite{guoTAC2021poly} returns 
a state-dependent matrix condition rather than an SOS condition. If such a 
state-dependent matrix condition can be solved at each time step along a trajectory 
of the system, then a control sequence that steers that trajectory to the origin is obtained. 
This idea is pursued in \cite{dai2021statedependent}. 

{\it Contribution.} We introduce a method to deal with the data-driven control 
design of nonlinear systems building up on and strengthening the results of 
\cite{de2019formulas} in several directions. 

We first consider nonlinear vector fields that are expressed 
as combinations of known nonlinear functions (not necessarily polynomials). 
We then derive conditions to design from data controllers that stabilize the 
closed-loop system via nonlinearity cancellations. This approach returns formulas 
for controller design which retain the same simplicity and compactness of the 
formulas established in \cite{de2019formulas} for linear systems, namely semi-definite 
programs (SDP) only depending on data. 

We then make the crucial observation that, were exact nonlinearity cancellation 
unfeasible, we can instead formulate an SDP that minimizes the
 norm of the matrix by which the nonlinearities enter the dynamics.  This idea is suggested 
 by a regularization procedure in which the hard constraint of the first approach, corresponding 
 to an exact nonlinearity cancellation, is lifted to an objective function, 
 corresponding to an {\em approximate} nonlinearity cancellation.
(In different contexts, this ``lifting" idea has been pursued 
 in \cite{bridging,breschi2021direct,dorfler2021certainty}). 
In general the design based on an approximate nonlinearity cancellation 
does not return globally stabilizing controllers, whence the need to explicitly 
characterize the region of attraction of the closed-loop system. We show that 
this is indeed possible by bounding the Lyapunov decrement via functions which 
are obtainable form data. We remark here that, although we focus on 
nonlinear discrete-time systems, analogous results can be derived for continuous-time systems too. 

To present the main ideas, we choose to give the results first for data that are not perturbed. 
The results are then extended to the case is which data are perturbed by process disturbances. 
In doing so, we show how our approach can  accomodate the presence of process disturbances 
not only during the collection of data used in the controller  design, but also during the execution 
of the control task and  provide estimates of \emph{robustly positively invariant sets} \cite{Blanchini1999}
for the closed-loop system.
The results are also extended to systems with nonlinearities 
that are not expressible as combination 
of known functions, thus significantly enlarging the class of nonlinear systems the 
approach can cope with. 

{\it Outline.} 
The framework is set in Section \ref{sec:framework}. The main results 
are discussed in Sections \ref{sec:exact} and \ref{sec:approx}, with some extensions 
in Section \ref{sec:ext}. Control design in the presence of disturbances 
and neglected nonlinearities is studied in Section \ref{sec:noise}. 
Some additional discussion is finally provided in Section \ref{sec:disc}.  

{\it Notation.} Throughout the paper, $\succ$ ($\succeq$) and $\prec$ ($\preceq$) 
denote positive and negative (semi)-definiteness, respectively; 
$\mathbb S^{n \times n}$ denotes the set of real-valued {symmetric} matrices
of dimension $n \times n$; $M^\top$ denotes the transpose of $M$.

\section{Framework} \label{sec:framework}

We consider a discrete-time system in the form
\begin{equation}
\label{system}
x^+ = A_\star Z_\star(x) + B u
\end{equation}  
where $x \in \mathbb R^{n}$ is the state and $u \in \mathbb R^{m}$ is the control input,
$A_\star \in \mathbb R^{n \times R}$, $B \in \mathbb R^{n \times m}$ 
are constant matrices, $Z_\star: \mathbb R^n \rightarrow \mathbb R^R$ is a
vector-valued function. Any nonlinear system 
in the form  $x^+ = f(x) + B u$ with $f: \mathbb R^n \rightarrow \mathbb R^n$ an arbitrary 
function can be written as in \eqref{system};
we adopt the representation \eqref{system} for convenience. 
In this paper, $A_\star$ and $B$ are regarded unknown while the 
following standing assumption is made for $Z_\star$.

\begin{assumption} \label{ass:Z}
We know a function $Z: \mathbb R^n \rightarrow \mathbb R^S$ such that any element of $Z_\star$
is also an element of $Z$. \quad $\Box$
\end{assumption} 

Under Assumption \ref{ass:Z}, system \eqref{system} reads equivalently as
\begin{equation}
\label{system2}
x^+ = A Z(x) + B u
\end{equation}  
with $A \in \mathbb R^{n \times S}$, and $A,B$ unknown.

Assumption \ref{ass:Z} amounts to considering systems
with known \emph{type} of dynamics (but possibly unknown parameters).
This assumption is satisfied in many practical cases such as
with mechanical and electrical systems where information about the
dynamics can be derived from first principles, but the
exact systems parameters may be unknown. We allow $Z$ to contain terms not present in $Z_\star$,
which may arise from an imprecise knowledge of the 
system dynamics. In this paper, we will directly consider the case 
where $Z$ contains \emph{both} linear and nonlinear functions, \emph{i.e.},
\begin{equation} \label{eq:Z}
Z(x) = \begin{bmatrix} x \\ \hline Q(x) \end{bmatrix}\,,
\end{equation}
with $Q: \mathbb R^n \rightarrow \mathbb R^{S-n}$ containing only nonlinear functions.
The special case where $Z(x)=x$ reduces the analysis to that of linear systems,
which have been the subject of numerous investigations,
as reviewed in the Introduction. 
In contrast, $Z(x)=Q(x)$ accounts for purely nonlinear systems, and just 
leads to simplified algorithms and results. We will exemplify this point in connection
with Theorem \ref{thm:exact}. 
Let 
\begin{equation} \label{dataset}
\mathbb D := \left\{ x(k),u(k) \right\}_{k=0}^T
\end{equation} 
be a dataset collected from the system with an experiment,
meaning that we have a set of state and input samples that 
satisfy $x(k+1)=AZ(x(k))+Bu(k)$ for $k=0,\ldots,T-1$, $T>0$.
The problem of interest is to determine, using $\mathbb D$, a control law $u=KZ(x)$ that stabilizes the system 
around the origin (globally or locally, both cases will be considered). 
Note that we might consider a control law $u=KH(x)$ with $H$ different from $Z$. 
As it will become clear soon, we focus on $u=KZ(x)$
as our approach is based on nonlinearity cancellation\,/\,minimization. 

The framework can be modified and/or extended in several directions: (i) Continuous-time systems
can be handled with similar arguments (Section \ref{subsec:CT});
(ii) The analysis extends to a more general class of nonlinear systems
(Section \ref{subsec:State-dependent input}); 
(iii) Noisy data and neglected nonlinearities are considered in Section \ref{sec:noise}.

\section{Exact nonlinearity cancellation}
\label{sec:exact}

We start by considering the scenario in which there exists a controller $K$
that linearizes the closed-loop dynamics, namely the scenario 
in which there exists a controller $K$
such that 
\begin{equation} \label{closed_ideal}
u=KZ(x) \quad \Longrightarrow \quad x^+ = M x
\end{equation}
for some matrix $M$ (which we will also require to be Schur\footnote{A matrix $M$ is said 
to be Schur if all its eigenvalues have
modulus less than one. For continuous-time systems, a matrix $M$ is said to be 
Hurwitz if all its eigenvalues have negative real part.}).

\subsection{Data-based closed-loop representation and 
control design for exact nonlinearity cancellation} \label{subsec:exact}

Consider the dataset  $\mathbb D$  in \eqref{dataset}, and define 
\begin{subequations}
\label{eq:data}
\begin{alignat}{4}
& U_0 := \begin{bmatrix} u(0) & u(1) & \cdots & u(T-1)  \end{bmatrix} 
\in \mathbb R^{m \times T} \,, \label{eq:data1} \\
& X_0 := \begin{bmatrix} x(0) & x(1) & \cdots & x(T-1)  \end{bmatrix} 
\in \mathbb R^{n \times T} \,, \label{eq:data2} \\
& X_1 := \begin{bmatrix} x(1) & x(2) & \cdots & x(T)  \end{bmatrix} 
\in \mathbb R^{n \times T} \,, \label{eq:data3} \\
& Z_0 := \begin{bmatrix} x(0) & x(1) & \cdots & x(T-1)  \\ 
Q(x(0)) & Q(x(1)) & \cdots & Q(x(T-1)) 
\end{bmatrix}  \in \mathbb R^{S \times T} \,, \nonumber \\ \label{eq:data4}
\end{alignat}
\end{subequations}
All the results of this paper rest on the following lemma. 
An analogous result was established in \cite[Lemma 1]{guo2020learning} 
for the case of polynomial systems. 
\begin{lem} \label{lem:main}
Consider any matrices $K \in \mathbb R^{m \times S}$,
$G \in \mathbb R^{T \times S}$ such that
\begin{equation} \label{eq:GK}
\begin{bmatrix} K \\ I_S \end{bmatrix} = \begin{bmatrix} U_0 \\ Z_0 \end{bmatrix} G \,.
\end{equation}
Let $G$ be partitioned as $G = \begin{bmatrix} G_1 & G_2 \end{bmatrix}$, where
$G_1 \in \mathbb R^{T \times n}$ and $G_2 \in \mathbb R^{T \times (S-n)}$.
Then, system \eqref{system} under the control law $u=KZ(x)$ results in
the closed-loop dynamics 
\begin{equation} \label{eq:GK_closed}
x^+ = Mx + N Q(x) 
\end{equation}
where $M:=X_1 G_1$ and $N:=X_1 G_2$.
\quad $\Box$
\end{lem} 

\emph{Proof.} 
The closed-loop dynamics resulting from the control law $u=KZ(x)$ is given by
\begin{subequations}
\label{eq:closed_loop_ideal}
\begin{alignat}{4}
x^+ &= \begin{bmatrix} B & A \end{bmatrix} \begin{bmatrix} K \\ I_S \end{bmatrix} Z(x) \\
&= \begin{bmatrix} B & A \end{bmatrix} \begin{bmatrix} U_0 \\ Z_0 \end{bmatrix} G Z(x) \,=\, X_1 G Z(x) \,.
\end{alignat}
\end{subequations}
The second identity follows from \eqref{eq:GK} while the last one
follows because the elements of $X_1,Z_0$ and $U_0$ satisfy the relation
$x(k+1) = A Z(x(k)) + Bu(k)$, $k=0,\ldots,T-1$, 
which, in compact form, gives $X_1 = AZ_0 +BU_0$.
\quad $\blacksquare$

Arrived at this stage, it is simple to derive a convex program
(specifically a semi-definite program (SDP))
that searches for a controller $K$ that cancels out the nonlinearities 
and renders the closed-loop system (globally) asymptotically stable.
Note that in next Theorem \ref{thm:exact} the decision variable 
$G_2$ represents the same quantity that appears in Lemma \ref{lem:main}.
The decision variables $Y_1,P_1$ are instead related to $G_1$ in 
Lemma \ref{lem:main} via $Y_1=G_1P_1$ with $P_1$ a positive definite matrix,
that is $Y_1$ defines a change of variable relative to $G_1$.
As it becomes clear from the proof of Theorem \ref{thm:exact}, this change of variable 
is instrumental to arrive at a convex formulation of the design program.

\begin{theorem} \label{thm:exact}
Consider a nonlinear system as in \eqref{system}, along with
the following SDP in the decision variables $P_1 \in \mathbb S^{n \times n}$,
$Y_1 \in \mathbb R^{T \times n}$, and
$G_2 \in \mathbb R^{T \times (S-n)}$:
\begin{subequations}
\label{eq:SDP}
\begin{alignat}{6}
& Z_0 Y_1 = \begin{bmatrix} P_1 \\ 0_{(S-n) \times n} \end{bmatrix} \,,
\label{eq:SDP1} \\
& \begin{bmatrix} P_1 & (X_1 Y_1)^\top  \\ 
X_1 Y_1 & P_1 \end{bmatrix}  \succ 0 \,, \label{eq:SDP2} \\
& Z_0 G_2 = \begin{bmatrix} 0_{n \times (S-n)} \\ I_{S-n} \end{bmatrix} \,,
\label{eq:SDP4} \\
& X_1 G_2 = 0_{n \times (S-n)} \,. \label{eq:SDP5} 
\end{alignat}
\end{subequations} 
If the SDP is feasible then the control law $u=KZ(x)$ with 
\begin{eqnarray} \label{eq:K_SDP}
K= U_0 \begin{bmatrix} Y_1 & G_2 \end{bmatrix}
\begin{bmatrix} P_1 & 0_{n \times (S-n)} \\
0_{(S-n) \times n} & I_{S-n}
\end{bmatrix}^{-1} 
\end{eqnarray}
linearizes the closed-loop dynamics, and renders the origin a globally
asymptotically stable equilibrium. \quad $\Box$
\end{theorem} 

\emph{Proof.} Suppose that \eqref{eq:SDP} is feasible. Let $G_1=Y_1P_1^{-1}$
and note that the two constraints \eqref{eq:SDP1} and \eqref{eq:SDP4} together 
yield 
\begin{equation} \label{eq:closed_solution}
Z_0 \begin{bmatrix} G_1 & G_2 \end{bmatrix} = I_S \,.
\end{equation}
This relation, combined with \eqref{eq:K_SDP}, gives
\begin{equation}
\begin{bmatrix} K \\ I_S \end{bmatrix} =
\begin{bmatrix} U_0 \\ Z_0 \end{bmatrix} 
\begin{bmatrix} G_1 & G_2 \end{bmatrix} ,
\end{equation}
which is \eqref{eq:GK}. By Lemma \ref{lem:main}, we conclude
that the closed-loop dynamics satisfies $x^+=Mx+NQ(x)$ with 
$M=X_1G_1$ and $N=X_1G_2$. By \eqref{eq:SDP5}, $N=0$. Hence,  
$K$ linearizes the closed-loop dynamics. Finally, note that 
\eqref{eq:SDP2} is equivalent to 
$P_1\succ 0$ and $(X_1Y_1)^\top P_1^{-1} (X_1Y_1) - P_1 \prec 0$. The latter, in turn,
is equivalent to $(X_1Y_1P_1^{-1})^\top P_1^{-1} (X_1Y_1P_1^{-1}) - P_1^{-1} \prec 0$.
By recalling that $Y_1P_1^{-1}=G_1$ and $X_1G_1=M$, we conclude that $M$
is Schur. 
(This also shows 
that $V(x)=x^\top P_1^{-1} x$ is a Lyapunov function for the closed-loop system.)
\quad $\blacksquare$
 
Theorem \ref{thm:exact} gives an extension to nonlinear
systems of the results in \cite{de2019formulas}. In fact, in the limit case where
$Z(x)=x$ we have $S=n$ and \eqref{eq:SDP} reduces to the first two constraints 
\eqref{eq:SDP1}-\eqref{eq:SDP2}, which appeared in \cite[Theorem 3]{de2019formulas}.
In general, \eqref{eq:SDP4}-\eqref{eq:SDP5} 
implement the linearization constraint, and \eqref{eq:SDP1}-\eqref{eq:SDP2} 
ensure a stable behavior for the linear dynamics. 
Note in particular that \eqref{eq:SDP4}, together with \eqref{eq:SDP1}, forms
a consistency relation which makes it possible to parametrize the 
closed-loop dynamics through data alone. 
The other extreme case occurs when $Z$ contains only nonlinear functions, 
\emph{i.e.,} when $Z(x)=Q(x)$. In this case, \eqref{eq:SDP} reduces to the 
two constraints \eqref{eq:SDP4}-\eqref{eq:SDP5}. This corresponds to 
a situation where the system has stable open-loop linear dynamics and the controller
is only responsible for canceling out all the nonlinearities. 

As a second remark, we observe that a necessary condition for the SDP \eqref{eq:SDP}
to be feasible is that $Z_0 $ has full row rank (this is indeed necessary 
to have both \eqref{eq:SDP1} and \eqref{eq:SDP4} fulfilled).
This requirement can be viewed as a condition on the \emph{richness} of the data, and is the 
natural generalization of the condition on the rank of $X_0$ that
appears in the linear case \cite[Theorem 3]{de2019formulas},
\cite[Theorem 16]{van2020data}. This condition
is weaker than having $[\begin{smallmatrix} U_0 \\ Z_0 \end{smallmatrix}]$
full row rank, which is instead necessary to identify $A,B$ from data, and this shows 
that learning a control law is in general {easier} than identifying the dynamics of the system.
Note that Lemma \ref{lem:main} indeed gives a data-based \emph{closed-loop} 
representation of the system dynamics, without any explicit estimate 
of the system matrices.

Having $[\begin{smallmatrix} U_0 \\ Z_0 \end{smallmatrix}]$ full row rank 
brings certain advantages, though. In fact, in this case, \emph{any}
controller that linearizes the closed-loop dynamics can be parametrized 
through the data. In particular, in this situation we obtain an  ``\emph{if and only if}" result,
meaning that \eqref{eq:SDP} is feasible and returns 
a stabilizing and linearizing controller whenever such a controller exists. We state the result 
but discuss it in Appendix \ref{subsec:param:exact} to maintain continuity. 

\begin{theorem} \label{thm:param:exact}
Suppose there exists a stabilizing and linearizing feedback controller,
\emph{i.e.}, a controller $K=[\begin{matrix} \overline K & 
\hat K \end{matrix}]$ such that 
\begin{subequations}
\label{eq:data}
\begin{alignat}{4}
& A+BK = \begin{bmatrix} \overline A + B \overline K & 0_{n \times (S-n)} \end{bmatrix} \\
& \overline A + B \overline K \,\text{ is Schur}
\end{alignat}
\end{subequations}
having partitioned $A=[\begin{matrix} \overline A & 
\hat A \end{matrix}]$ with $\overline A \in \mathbb R^{n \times n}$.
Let $[\begin{smallmatrix} U_0 \\ Z_0 \end{smallmatrix}]$ have full row rank. 
Then \eqref{eq:SDP} is feasible and $K$ can be written as in
\eqref{eq:K_SDP}  for some $Y_1,P_1,G_2$ satisfying \eqref{eq:SDP}. \quad $\Box$
\end{theorem}  

\textbf{Example 1.} Consider the Euler discretization of an inverted pendulum  
\begin{subequations} 
\label{eq:sys_example_pendulum}
\begin{alignat}{2}
& x_1^+ = x_1 + T_s x_2 \\   
& x_2^+ = \displaystyle  \frac{T_s  g}{\ell} \sin x_1 + \left( 1 - \frac{T_s  \mu}{m \ell^2} \right) x_2 +
\frac{T_s }{m \ell^2}u \,,
\end{alignat}
\end{subequations}
where $T_s$ is the sampling time, $m$ is the mass to be balanced, 
$\ell$ is the distance from the base to the center of mass of the balanced body,
$\mu$ is the coefficient of rotational friction, and $g$ is the acceleration
due to gravity. The states $x_1, x_2$ are the angular position and velocity, respectively, 
$u$ is the applied torque. The system has an unstable equilibrium in 
$(x, u)=(0,0)$, corresponding to the pendulum upright position, which we want to stabilize.
Suppose that the parameters are $T_s= 0.1$, $m=1$, $\ell=1$, $g=9.8$ and $\mu=0.01$.

We choose $Z(x)=\begin{bmatrix} x_1 & x_2 & \sin(x_1) \end{bmatrix}^\top$,
and regard all the parameters $T_s,m,\ell,g,\mu$ as unknown
(here, a correct choice for $Z(x)$ simply derives from physical considerations,
namely Lagrange's equations of motion). We collect data by running an experiment with input uniformly 
distributed in $[-0.5,0.5]$, and with an initial state within the same interval. 
We collect $T=10$ samples (corresponding to the motion of 
the pendulum that oscillates around the upright position).
The SDP \eqref{eq:SDP}  is feasible and we obtain 
$K = \begin{bmatrix} -23.5641  & -10.3901 &  -9.8 \end{bmatrix}$.
The resulting control law indeed cancels out the nonlinearity ensuring global asymptotic stability. 
\quad $\blacksquare$ 

\textbf{Example 2.} Consider the polynomial system
\begin{subequations} 
\label{eq:sys_example1}
\begin{alignat}{2}
& x_1^+ =  x_2 + x_1^3 + u \\   
& x_2^+ = 0.5 x_1 \,.
\end{alignat}
\end{subequations}
Suppose that we choose 
\begin{equation}
Z(x) = \begin{bmatrix} \label{eq:Z_example1}
x^\top & x_1^2 & x_2^2 & x_1x_2 &
x_1^3 & x_2^3 & x_1 x_2^2 & x_1^2 x_2
\end{bmatrix}^\top,
\end{equation}
\emph{i.e.,} we capture the nonlinearity by including all the 
possible monomials up to degree $3$.
The equilibrium of the unforced system ($u=0$) is only locally 
asymptotically stable (\emph{e.g.}, any initial condition such that 
$x_1(0)>1$ and $x_2(0)\ge 0$ 
leads to a divergent solution). 
We collect data by running an experiment with input uniformly 
distributed in $[-0.5,0.5]$, and with an initial state within the same interval.
We collect $T=10$ samples. The SDP is feasible and returns the 
controller \smallskip

\small
\begin{equation} \label{eq:K_example1}
K=  \big[
\underbrace{0}_{x_1} \,\,\,  \underbrace{-1.0007}_{x_2}  
\,\,\,  \underbrace{0}_{x_1^2} \,\,\,
\underbrace{0}_{x_2^2} \,\,\, \underbrace{0}_{x_1 x_2} 
\,\,\, \underbrace{-1}_{x_1^3} \,\,\,
\underbrace{0}_{x_2^3} \,\,\, \underbrace{0}_{x_1 x_2^2} 
\,\,\, \underbrace{0}_{x_1^2 x_2} \big]
\end{equation}

\normalsize
The SDP correctly assigns the value $-1$ to the sixth entry of $K$,
and automatically discovers that no other nonlinearities are present.
The resulting control law is $u=-1.0007 x_2 - x_1^3$ 
and ensures global asymptotic stability. \quad $\blacksquare$


The examples show that even a few samples may suffice to learn a stabilizing control policy.
In fact, in terms of number of data points, the only necessary condition in \eqref{eq:SDP} comes from
having $Z_0 $ full row rank, and this condition can be met even with $T=S$ samples.
The situation may be different with noisy data as we discuss in Section \ref{sec:noise}.
As a second remark, note that this approach differs from the approach 
in \cite{de2019formulas}, which considers \emph{linear} control laws. 
This new approach considers \emph{nonlinear} control laws; this
is indeed essential to achieve nonlinearity cancellation (or nonlinearity minimization, 
if cancellation is impossible, as we discuss in Section \ref{sec:approx}).

\subsection{Nonlinearity cancellation as a minimization problem}
\label{subsec:lifting}

A variant of \eqref{eq:SDP} consists in approaching the 
design problem as a \emph{minimization} problem, namely as the problem 
of finding a controller that {minimizes} the nonlinearity in closed loop 
with respect to some chosen norm. 
  
\begin{theorem} \label{thm:exact_lift}
Consider a nonlinear system as in \eqref{system}
along with the following SDP in the decision variables $P_1 \in \mathbb S^{n \times n}$,
$Y_1 \in \mathbb R^{T \times n}$, and
$G_2 \in \mathbb R^{T \times (S-n)}$:
\begin{subequations}
\label{eq:2SDP}
\begin{alignat}{2}
\textrm{minimize}_{P_1,Y_1,G_2} \quad
& \|X_1 G_2\| \label{eq:2SDPa} \\ 
\textrm{subject to} \quad 
& Z_0 Y_1 = \begin{bmatrix} P_1 \\ 0_{(S-n) \times n} \end{bmatrix} \,,
\label{eq:2SDP1} \\
& \begin{bmatrix} P_1 & (X_1 Y_1)^\top  \\ 
X_1 Y_1 & P_1 \end{bmatrix}  \succ 0 \,, \label{eq:2SDP2} \\
& Z_0 G_2 = \begin{bmatrix} 0_{n \times (S-n)} \\ I_{S-n} \end{bmatrix} \,.
\label{eq:2SDP4} 
\end{alignat}
\end{subequations}
If this SDP is feasible and the solution achieves zero cost 
(\emph{i.e.}, $\|X_1 G_2\|=0$) then the control law $u=KZ(x)$ with 
$K$ given by \eqref{eq:K_SDP}
linearizes the closed-loop dynamics, and renders the origin a globally
asymptotically stable equilibrium. 

(Here, $\|\cdot\|$ is any norm.)
\quad $\Box$
\end{theorem} 

\emph{Proof}. The proof is analogous to the proof of Theorem \ref{thm:exact} and therefore
omitted. \quad $\blacksquare$

\textbf{Example 3.} Consider again system
\eqref{eq:sys_example1} under the same experimental setting as before. 
The SDP \eqref{eq:2SDP} is feasible and we obtain (we use the induced $2$-norm 
in \eqref{eq:2SDPa})
\begin{equation} \label{eq:K_example2}
K= \begin{bmatrix} 
0.0001&  -1.0007 &  0  &  0  &  0 &  -1 &
0  &  0  & 0
\end{bmatrix} 
\end{equation}
As before, the program correctly assigns the value $-1$ to the sixth entry of $K$.
Note that when nonlinearity cancellation is possible, \eqref{eq:SDP}
and \eqref{eq:2SDP} are \emph{equivalent} in the sense that their feasible sets coincide.
The controller in \eqref{eq:K_example2} 
differs from the one in \eqref{eq:K_example1} simply because
there are infinitely many stabilizing and linearizing controllers and neither \eqref{eq:SDP}
nor \eqref{eq:2SDP} involve constraints other than stability and linearization. \quad $\blacksquare$

\section{Approximate nonlinearity cancellation} \label{sec:approx}
 
\subsection{Control design for approximate nonlinearity cancellation}
\label{subsec:approx}

There is a simple yet important difference between 
\eqref{eq:SDP} and its lifted version \eqref{eq:2SDP}. The difference 
is that the latter is always feasible when the former is feasible 
and this implies that we can always use \eqref{eq:2SDP} in place of \eqref{eq:SDP}
when exact nonlinearity cancellation is possible. However, 
\eqref{eq:2SDP} can be adopted even when exact cancellation is impossible,
in which case \eqref{eq:SDP} is instead infeasible.

The next result indeed addresses the scenario where 
exact cancellation is impossible. It shows in particular that, in this case,
we can still have stability guarantees.

\begin{theorem} \label{thm:approx}
Consider a nonlinear system as in \eqref{system}, along with the SDP
\eqref{eq:2SDP}. Assume that 
\begin{equation} \label{eq:limitQx}
\lim_{|x|\to 0} \frac{|Q(x)|}{|x|}=0 \,.
\end{equation}
If the SDP is feasible then $u=KZ(x)$, with 
$K$ as in \eqref{eq:K_SDP}, renders the origin an asymptotically stable equilibrium. 
\quad $\Box$
\end{theorem}

\emph{Proof.} The first part of the proof is analogous to that of Theorem \ref{thm:exact}.
Suppose that \eqref{eq:2SDP} is feasible. Let $G_1=Y_1P_1^{-1}$,
and note that the two constraints \eqref{eq:2SDP1} and \eqref{eq:2SDP4} together 
yield $Z_0 \begin{bmatrix} G_1 & G_2 \end{bmatrix} = I_S$.
This identity, along with \eqref{eq:K_SDP}, gives
\eqref{eq:GK}. By Lemma \ref{lem:main}, we have
that the closed-loop dynamics satisfies $x^+=Mx+NQ(x)$, where 
$M=X_1G_1$ and $N=X_1G_2$. Although $N$ might be different from zero,
\eqref{eq:2SDP2} ensures that $M$ is Schur. 
Asymptotic stability thus follows from \eqref{eq:limitQx}.
\quad $\blacksquare$

In Theorem \ref{thm:approx}, the condition $\lim_{|x|\to 0} \frac{|Q(x)|}{|x|}=0$ 
ensures that the linear dynamics dominates the nonlinear dynamics around the origin.
In turn, as shown in the next subsection, this guarantees that we can obtain an estimate of the 
\emph{region of attraction}. This condition is satisfied for many systems of practical relevance,
for instance is satisfied by \emph{any} {polynomial} system. More generally, 
the condition $\lim_{|x|\to 0} \frac{|Q(x)|}{|x|}=0$ can be rephrased by 
asking that $Z$ is differentiable at $x=0$ and satisfies $Z(0)=0$. In fact, in this case 
$Q$ is differentiable at $x=0$ and satisfies $Q(0)=0$, hence it
admits a Taylor's expansion at $x=0$, namely we have
\begin{subequations}
\begin{alignat}{4}
Q(x) & = \left[\frac{\partial Q}{\partial x }\right]_{x =0} x + r(x)
\end{alignat} 
\end{subequations}
with $r: \mathbb R^n \rightarrow \mathbb R^{S-n}$ a differentiable function of the state such that  
$\lim_{|x|\to 0} \frac{|r(x)|}{|x|}=0$. Thus, system \eqref{system} can be equivalently represented as 
\begin{subequations}
\begin{alignat}{4}
x^+ 
& = \overline A x + \hat A Q(x) + Bu \label{24b}\\
& = (\overline A + \hat A F)  x + \hat A r(x) + Bu \label{24c}
\end{alignat} 
\end{subequations}
where we have partitioned $A$ as $A = \begin{bmatrix}\, \overline A & \hat A  \end{bmatrix}$
with $\overline A \in \mathbb R^{n \times n}$. 
Hence, Theorem \ref{thm:approx} becomes applicable with 
$Q$ replaced by $r$,
where $r$ can be determined from $Q$.
As an example, for the inverted pendulum this reasoning leads to
$r(x)=\sin(x_1)-x_1$,
which gives $\lim_{|x|\to 0} \frac{|r(x)|}{|x|}=0$ 
(for the inverted pendulum
Theorem \ref{thm:approx} reduces in any case to Theorem \ref{thm:exact_lift} because
exact cancellation is possible). 

%

We point out that there exists a counterpart of Theorem \ref{thm:param:exact},
which provides conditions under which we can parametrize 
\emph{all} feedback controllers that ensure local stability through 
a stable linear dynamics. As before, we state the result 
but prove it in the appendix (Appendix \ref{subsec:param:approx}) to maintain continuity. 

\begin{theorem} \label{thm:param:approx}
Suppose that there exists a feedback controller,
$K=[\begin{matrix} \overline K & 
\hat K \end{matrix}]$ such that $\overline A + B \overline K$ is Schur,
having partitioned $A=[\begin{matrix} \overline A & 
\hat A \end{matrix}]$ with $\overline A \in \mathbb R^{n \times n}$.
Let $[\begin{smallmatrix} U_0 \\ Z_0 \end{smallmatrix}]$ have full row rank. 
Then \eqref{eq:2SDP} is feasible and $K$ can be written as in
\eqref{eq:K_SDP} for some $P_1,Y_1,G_2$ satisfying \eqref{eq:2SDP}. \quad $\Box$
\end{theorem}  

\subsection{Estimating the region of attraction} \label{sec:ROA}

\begin{definition} \label{def:ROA}
A set $\mathcal S$ is called \emph{positively invariant} (PI) for the system $x^+ = f(x)$ 
if for every $x(0) \in \mathcal S$ the solution is such that $x(t) \in \mathcal S$ for $t>0$.
Let $\overline x$ be an asymptotically stable equilibrium point for the system $x^+ = f(x)$.
A set $\mathcal R$ defines a \emph{region of attraction}  (ROA) for the system relative to $\overline x$
if for every $x(0) \in \mathcal R$ we have $\lim_{t \rightarrow \infty} 
x(t) = \overline x$. \quad $\Box$
\end{definition}

Building on Theorem \ref{thm:approx}, we can give estimates of the ROA 
for the closed-loop system relative to the equilibrium $\overline x=0$. Consider the same
conditions as in Theorem \ref{thm:approx} and note
that $V(x):=x^\top P_1^{-1} x$ is 
a Lyapunov function for the linear part of the dynamics. In particular, 
\begin{eqnarray} \label{eq:Lyap}
&& \hspace{-0.7cm}  V(x^+) - V(x) = \nonumber \\ 
 && \qquad  \hspace{-0.7cm} 
 \underbrace{(Mx+NQ(x))^\top P_1^{-1} (Mx+NQ(x)) - x^\top P_1^{-1} x}_{=: h(x)} . \nonumber \\ 
\end{eqnarray}
where the matrices $M,N$ and $P_1$ are all computable from data.
We immediately obtain the following result.

\begin{proposition} \label{prop:RoA}
Consider the same setting as in Theorem \ref{thm:approx}.
Let $\mathcal V := \{x: h(x) < 0 \}$ with $h(x)$ as in \eqref{eq:Lyap},
and consider the Lyapunov function $V(x)=x^\top P_1^{-1} x$. Then, any sub-level 
set $\mathcal R_\gamma := \{ x: V(x) \leq \gamma \}$ of $V$ contained in $\mathcal V \cup \{0\}$ 
is a PI set for the closed-loop system
and defines an estimate of the ROA relative to $\overline x=0$. \quad $\Box$
\end{proposition}

We close this section with an example that illustrates both
Theorem \ref{thm:approx} and Proposition \ref{prop:RoA}.

\textbf{Example 4.} Consider the nonlinear system
\begin{subequations}\label{exmp:5}
\begin{alignat}{2}
& x_1^+ =  x_2 + x_1^3 + u \\   
& x_2^+ = 0.5 x_1 + 0.2 x_2^2
\end{alignat}
\end{subequations}%
\begin{figure*}[!t]
\normalsize
\setcounter{mytempeqncnt}{\value{equation}}
\setcounter{equation}{27}
\begin{equation}
\label{eq:K_example3}
K=  \big[
\underbrace{-0.0113}_{x_1} \,\,\,  \underbrace{-1.0862}_{x_2}  \,\,\,  
\underbrace{0.0005}_{x_1^2} \,\,\,
\underbrace{0}_{x_2^2} \,\,\, \underbrace{0.0039}_{x_1 x_2} 
\,\,\, \underbrace{-1.0010}_{x_1^3} \,\,\,
\underbrace{-0.0130}_{x_2^3} \,\,\, \underbrace{0.0119}_{x_1 x_2^2} \,\,\, 
\underbrace{-0.0010}_{x_1^2 x_2} \big]        
\end{equation}
\begin{equation}
\label{eq:MN_example2}
M = \begin{bmatrix} -0.0113 &  -0.0862 \\ 0.5000 & 0 \end{bmatrix}, \quad
N = \begin{bmatrix} 
0.0005 & 0 & 0.0039 & -0.0010 & -0.0130 & 0.0119 & -0.0010 \\
0 & 0.2000 & 0 & 0 & 0 & 0 & 0 \end{bmatrix}
\end{equation}
\caption{Results for Example 4. Left: Sets $\mathcal V$ and 
$\mathcal R_\gamma$
in grey and black color, respectively, for the controller $K$ in \eqref{eq:K_example3}
(we recall that $\mathcal R_\gamma$ is a valid estimate for the ROA);
Middle: Sets $\mathcal V$ and $\mathcal R_\gamma$
in grey and black color, respectively, for the controller $K$ in \eqref{eq:K1_example3};
Right: Sets $\mathcal V$, $\mathcal R_\gamma$ and $\mathcal R$ (exact ROA) 
for the controller $K$ in \eqref{eq:K1_example3}. The set $\mathcal R$ is displayed in 
red color.} \label{fig:ex3}
{\includegraphics[width=6.2cm]{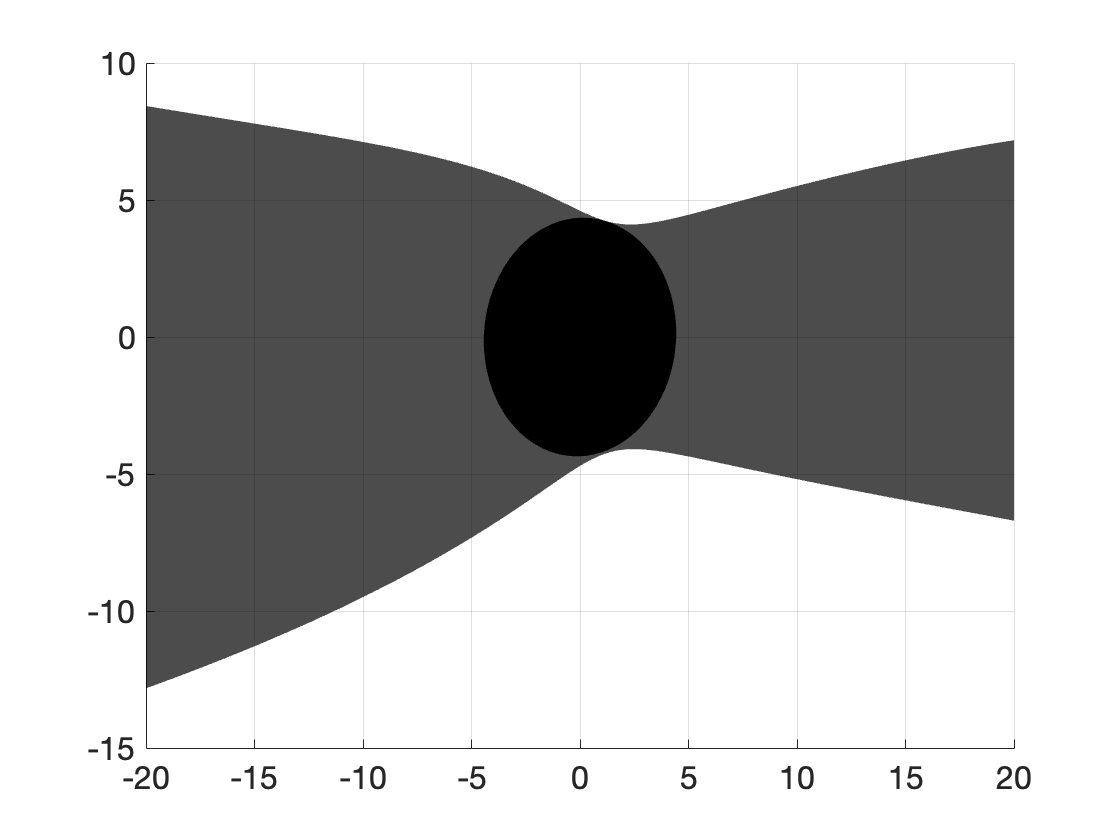}} \hspace*{-0.5cm}
{\includegraphics[width=6.2cm]{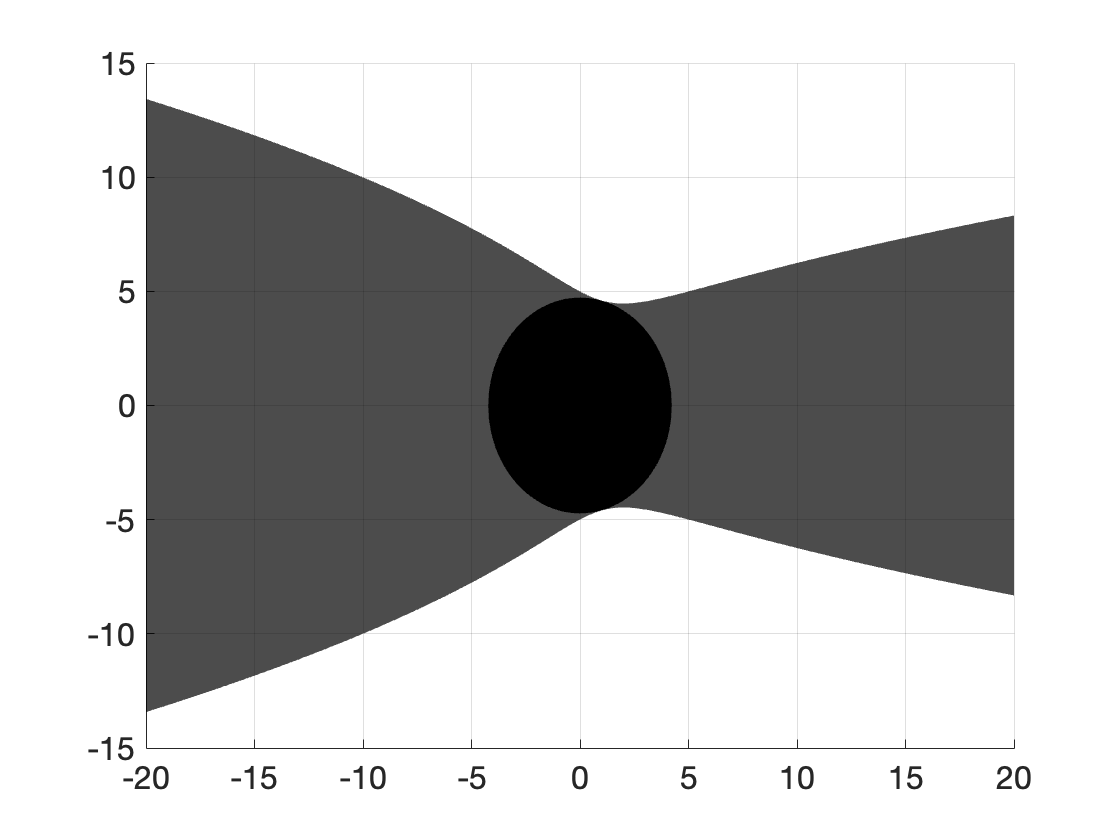}} \hspace*{-0.5cm} 
{\includegraphics[width=6.2cm]{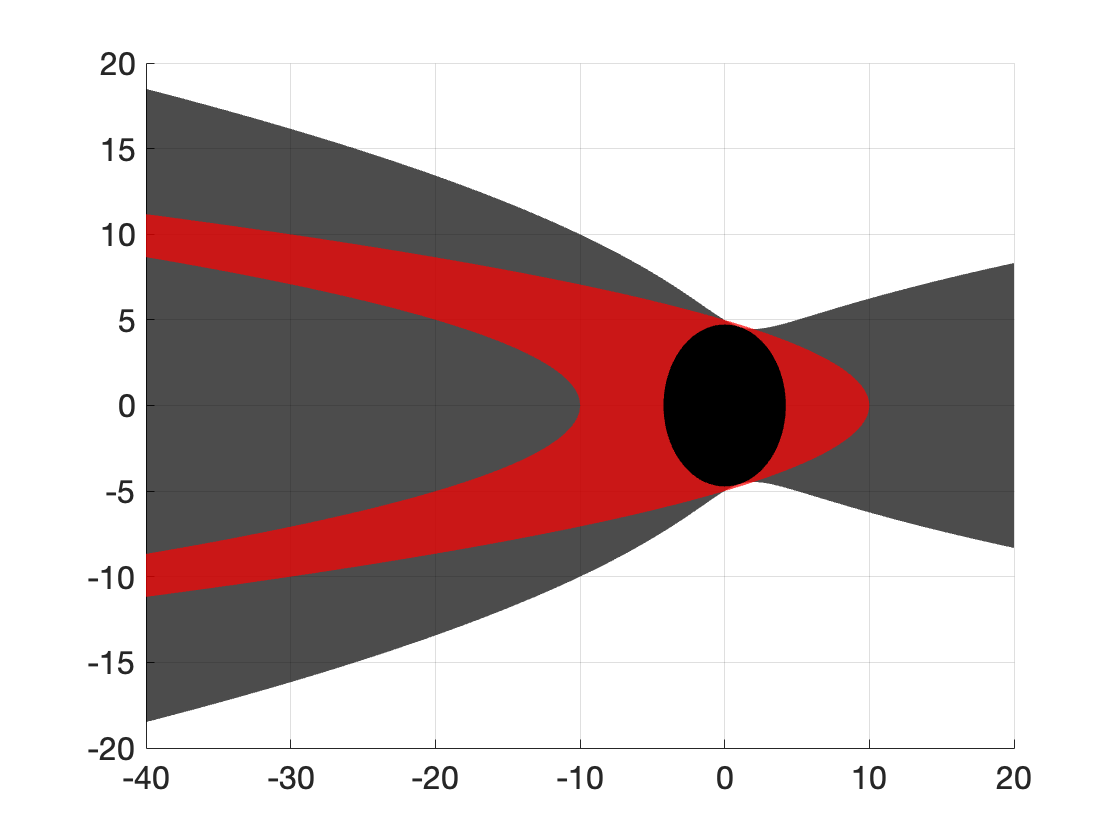}} 
\setcounter{equation}{\value{equation}}
\end{figure*}%
under the same experimental setting as before, in particular 
$Z(x)$ is as in \eqref{eq:Z_example1}. Exact nonlinearity
cancellation is now impossible. Nonetheless, 
the SDP \eqref{eq:2SDP} is feasible and returns the controller 
$K$ in \eqref{eq:K_example3} at the top of the page
(we take the induced $2$-norm in the objective function). 
For this controller, we numerically determine the set $\mathcal V = \{x: h(x) < 0\}$ over
which the Lyapunov function $V(x)=x^\top P_1^{-1} x$ decreases
and a sub-level set $\mathcal R_\gamma$ of $V$
contained in $\mathcal V \cup \{0\}$ which gives a valid estimate of the ROA.
These two sets are displayed in Figure \ref{fig:ex3} (Left). 
We note that the SDP \eqref{eq:2SDP} 
almost assigns the value $-1$ to the sixth entry of $K$, thus reducing the
effect of the nonlinearity on the first state component. Specifically, this controller 
results in the matrices $M$ and $N$ reported in \eqref{eq:MN_example2},
and the matrix $N$ has indeed minimum norm $\|N\|=0.2$ 
(this value cannot be further reduced because the term $0.2x_2^2$ cannot be canceled out). 

The approach that we just described for estimating the ROA is fully automatic
and is generically applicable. Note, however, that once we compute a controller $K$
then we can pursue \emph{any} approach (data- or model-based) to estimate 
the ROA. In fact, the SPD \eqref{eq:2SDP} returns the \emph{exact} description of 
the closed-loop dynamics: $x^+ = \begin{bmatrix} M&N \end{bmatrix} Z(x)$
(we stress that this expression does not correspond to identifying open-loop dynamics
of the system). 
From this description, we can then indeed apply any technique 
to find Lyapunov functions and estimate the ROA, 
see for instance \cite[Section 8.2]{KHALIL2002}.

%
%
%
%
%
%
%

To illustrate this point in a simple manner, suppose that \eqref{eq:2SDP} returns
\begin{equation} \label{eq:K1_example3}
K = \big[
\underbrace{0}_{x_1} \,\,\,  \underbrace{-1}_{x_2}  \,\,\,  \underbrace{0}_{x_1^2} \,\,\,
\underbrace{0}_{x_2^2} \,\,\, \underbrace{0}_{x_1 x_2} \,\,\, \underbrace{-1}_{x_1^3} \,\,\,
\underbrace{0}_{x_2^2} \,\,\, \underbrace{0}_{x_1 x_2^2} \,\,\, \underbrace{0}_{x_1^2 x_2} \big]       
\end{equation}
(this is indeed what we obtain with a variant of \eqref{eq:2SDP}, 
see next \eqref{eq:3SDP}), from which we have
\begin{equation*} 
M= \left[ \begin{array}{cc} 0 & 0 \\ 0.5 & 0 \end{array} \right], \quad
N= \left[ \begin{array}{cc|c}
0 & 0 & 0_{1 \times 5} \\ 0.2 & 0 & 0_{1 \times 5} \end{array} \right],
\end{equation*}
or, equivalently, 
\begin{subequations}
	\label{exact.RoA}
\begin{alignat}{2}
& x_1^+ =  0 \\   
& x_2^+ = 0.5 x_1 + 0.2 x_2^2 \,.
\end{alignat}
\end{subequations}
From the closed-loop dynamics we conclude that the exact ROA
is given by the set $\mathcal R := \{x: |0.5 x_1 + 0.2 x_2^2| < 5 \}$. 
In fact, the solution to system \eqref{exact.RoA} is given by   $x_1(t)=0$ for  $t\ge 1$ e 
$x_2(t)= b^{-1} (b(ax_1(0)+ b x_2(0)^2))^{2^{t-1}}$ for $t\ge 2$, 
with $a=0.5$ and $b=0.2$. Hence, the solution converges asymptotically 
if and only if $|b(ax_1(0)+ b x_2(0)^2)|<1$, from which one infers 
the ROA $\mathcal R$ specified above. 
This is a situation where it is simple to exactly compute by inspection the ROA,
which gives a better result with respect to the automatic
procedure, \emph{cf.} Figure \ref{fig:ex3} (Middle, Right). 
The automatic procedure, however, is applicable even when an exact 
description of the closed-loop dynamics is not available, as it is the case 
when noisy data are being measured, a case examined in Section \ref{sec:noise}. 
\quad $\blacksquare$

We conclude this section with a few additional remarks. 

As a first comment, note that the SDP \eqref{eq:2SDP} can also be
used to infer the stability properties of any controller $K$ for which 
a solution to \eqref{eq:GK} exists. This can be done by regarding 
\eqref{eq:K_SDP} as an additional constraint to \eqref{eq:2SDP}, \emph{i.e.},
by adding the constraint 
\begin{eqnarray*} 
U_0 \begin{bmatrix} Y_1 & G_2 \end{bmatrix} = 
K \begin{bmatrix} P_1 & 0_{n \times (S-n)} \\
0_{(S-n) \times n} & I_{S-n}
\end{bmatrix} 
\end{eqnarray*}
which is convex. This can be useful whenever a controller is inferred based 
on physical intuition and we want to determine closed-loop stability properties 
\emph{before} inserting the controller into the loop. For the same reason, by adding the constraint 
$U_0 \begin{bmatrix} Y_1 & G_2 \end{bmatrix}=0$ we infer the ROA for the open-loop system. 

As a final observation, we mention a particularly effective variant of \eqref{eq:2SDP}:
\begin{subequations}
\label{eq:3SDP}
\begin{alignat}{2}
\textrm{minimize}_{P_1,Y_1,G_2,X,V} \quad
& \text {trace} (X) + \text {trace} (V) \\ 
\textrm{subject to} \quad 
& \eqref{eq:2SDP1}-\eqref{eq:2SDP4} \\
& \begin{bmatrix} X & X_1 G_2  \\ 
(X_1 G_2)^\top & V \end{bmatrix}  \succeq 0 \label{eq:3SDP1} \,.
\end{alignat}
\end{subequations}
This SDP uses the trace as a convex envelope of the rank \cite{Fazel2001}, hence it
searches for solutions yielding a \emph{sparse} nonlinear term $N=X_1 G_2$, which 
can be useful to analyse properties of the closed-loop system, including the ROA. 
Applied to Example 4, this SDP indeed systematically returns a controller with 
third-to-ninth entries as in \eqref{eq:K1_example3}. If we further regularize 
\eqref{eq:3SDP} by enforcing a sparsity term for $X_1 Y_1$, the SDP exactly returns \eqref{eq:K1_example3}
(systematically for different datasets).
In a sense, the cost function in \eqref{eq:3SDP} is analogous to regularization terms
used in regression algorithms to penalize {complex} models \cite{Smola2001}. 
The difference is that 
\eqref{eq:3SDP} promotes low-complexity (sparse) \emph{closed-loop} systems (the matrix $X_1G_2$), and this
favours low-complexity (sparse) control laws.


 \section{Extensions}
\label{sec:ext}

The proposed approach can be extended in many directions. 
In this section, we discuss two of them.

\subsection{Continuous-time systems}
\label{subsec:CT}

Continuous-time systems can be treated in a similar way to the discrete-time
case, we will report the main differences. Suppose that we have a continuous-time 
system 
\begin{equation}
\label{systemTC}
\dot x = A Z(x) + B u
\end{equation}   
and that we make an experiment on it. Sampling the observed 
trajectory with sampling time $T_s>0$ we collect data matrices
$U_0,X_0,Z_0 ,X_1$ with $U_0,X_0$ and $Z_0 $ as in \eqref{eq:data1}, \eqref{eq:data2} 
and \eqref{eq:data4}, respectively, and with
$X_{1} := [\dot x(0) \,\,\, \dot x(T_s) \,\, \cdots \,\,  \dot x((T-1)T_s)]$.
It is readily seen that these data matrices satisfy the relation
$X_1=AZ_0 +BU_0$. As a consequence, the same analysis 
carried out in Section \ref{sec:exact} and \ref{sec:approx} carries over to 
the present case. The only modification occurs in the Lyapunov stability 
condition which reads $X_1 Y_1+ (X_1 Y_1)^\top \prec 0$ instead of 
\eqref{eq:2SDP2} (or \eqref{eq:SDP2}). In fact, recalling that the matrix $M$ 
that dictates the linear dynamics in closed loop is given by
$M=X_1 Y_1P_1^{-1}$,
the above Lyapunov inequality gives $P_1^{-1} M + M^\top P_1^{-1} \prec 0$, and this
implies that $M$ is Hurwitz (with Lyapunov function $V(x)=x^\top P_1^{-1}x$).
Hence, \eqref{eq:2SDP} (\eqref{eq:SDP} is analogous) becomes
\begin{subequations}
\label{eq:CT_SDP}
\begin{alignat}{2}
\textrm{minimize}_{P_1,Y_1,G_2} \quad
& \| X_1G_2 \| \\ 
\textrm{subject to} \quad 
& \eqref{eq:2SDP1},\eqref{eq:2SDP4} \\
& X_1 Y_1+ (X_1 Y_1)^\top \prec 0 \,,
\end{alignat}
\end{subequations}
and the (continuous-time) control law is given by $u=KZ(x)$ with $K$ as in \eqref{eq:K_SDP}. 

For estimating the ROA we can proceed as in Section \ref{sec:ROA},
we omit the details since they are straightforward. 

\subsection{A more general class of nonlinear systems}
\label{subsec:State-dependent input}

We now turn our attention to the case of systems 
\begin{equation}
\label{system.g(x).more.gen}
x^+ = A_\star \mathcal{Z}_\star(\xi)
\end{equation}  
where $\xi :=[\begin{smallmatrix} x \\ u \end{smallmatrix}]$, 
$A_\star \in \mathbb R^{n \times R}$ is an unknown constant matrix and 
where $\mathcal{Z}_\star: \mathbb R^{n+m}  \rightarrow \mathbb R^{R}$ is a 
vector-valued function of the state and the input. 
System \eqref{system.g(x).more.gen} is more general than \eqref{system} 
for it allows \emph{both} the state $x$ and the input $u$ to enter the dynamics nonlinearly. 
We rephrase  Assumption \ref{ass:Z} as follows: 

\begin{assumption} \label{ass:Z-W.more.gen}
We know a function $\mathcal{Z}: \mathbb R^{n+m} \rightarrow \mathbb R^{S}$ 
such that any element of $\mathcal{Z}_\star$ is also an element of $\mathcal{Z}$. \quad $\Box$
\end{assumption} 

Under this assumption, \eqref{system.g(x).more.gen} can be equivalently written as 
$x^+ = A \mathcal{Z} (\xi)$
with $A\in \R^{n\times S}$ an unknown matrix. As before, we allow 
$\mathcal{Z}(\xi)$ to contain both $\xi$ and the nonlinear function 
$\mathcal{Q}:\R^{n+m} \to \R^{S-n-m}$, namely we consider
\begin{equation}\label{Z(x,u)}
\mathcal{Z}(\xi)= \begin{bmatrix} \xi \\ \mathcal{Q}(\xi) \end{bmatrix}.
\end{equation}
The presence of $\mathcal{Q}(\xi)$ makes it difficult to adopt a similar design as in the previous sections, 
unless one regards the control input $u$ as a state variable and extends the dynamics to include 
the controller dynamics. 
This ``adding one integrator" tool, which has been widely used in control theory, reduces the 
design of the controller for \eqref{system.g(x).more.gen} to the case with constant 
input vector fields previously studied, as we detail below.  
 
Let us add the controller dynamics in the form $u^+ = v$,
with $v\in \mathbb{R}^m$ a new control input.
This extension leads to the system 
\begin{equation}\label{nonl.new.form.more.gen}
\xi^+ = \mathcal{A} \mathcal{Z}(\xi)+\mathcal{B} v,
\end{equation}
where 
\begin{equation}
\mathcal{A} := \begin{bmatrix} \, \overline A & \hat A \\
{0}_{m\times (n+m)} & {0}_{m\times (S-n-m)} \\ \end{bmatrix}, \quad
\mathcal{B} := \begin{bmatrix} 0_{n\times m} \\ I_m \end{bmatrix}
\end{equation}
having partitioned $A$ as $A=\begin{bmatrix}\, \overline A & \hat A  \end{bmatrix}$ with 
$\overline A\in \R^{n\times (n+m)}$. 
We therefore arrived at a representation which allows us to proceed as in 
the previous sections.
We collect the dataset $\{x(k), u(k), v(k)\}_{k=0}^T$ from the system 
and define the data matrices 
\begin{equation*}
\begin{array}{rl}
V_0 :=  & \hspace{-3mm}   \begin{bmatrix}
v(0)& v(1) & \ldots & v(T-1)
\end{bmatrix}\in \R^{m\times T} \\[0.2cm]
\Xi_0 := & \hspace{-3mm}   \begin{bmatrix} \xi(0) & \xi(1) & \ldots & \xi(T-1) 
\end{bmatrix}\in \R^{(n+m)\times T}  \\[0.2cm]
\Xi_1 :=  & \hspace{-3mm}   \begin{bmatrix} \xi(1) & \xi(2) & \ldots & \xi(T) 
\end{bmatrix}\in \R^{(n+m)\times T} \\[0.2cm]
{\footnotesize \mathcal{Z}_0 :=} & \hspace{-3mm}  
{\footnotesize \begin{bmatrix} \xi(0) & \ldots & \xi(T-1) \\[0.5mm]
Q(\xi(0)) & \ldots & Q(\xi(T-1))\\[0.5mm]
\end{bmatrix}\hspace{-1.5mm} 	\in \R^{S\times T}}
\end{array}
\end{equation*}
which satisfy the identity $\Xi_1 = \mathcal{A} Z_0 + \mathcal{B} V_0$. 
	
The following result parallels Theorem \ref{thm:approx}.
\begin{corollary}\label{cor:approx.extended.more.gen}
Consider a nonlinear system as in  \eqref{system.g(x).more.gen}, and assume that 
$\lim_{|\xi|\to 0} \frac{| \mathcal{Q}(\xi)|}{ |\xi |} =0$. 
Consider the following  SDP in the decision variables 
$Y_1\in \mathbb{R}^{T\times (n+m)}$,  $G_2\in \mathbb{R}^{T\times (S-n-m)}$,
$P_1\in \mathbb{S}^{(n+m)\times (n+m)}$:
\begin{subequations}
\label{eq:2SDP.extended.more.gen}
\begin{alignat}{2}
\textrm{minimize}_{P_1, Y_1, G_2} \quad
&  \| \Xi_1 G_2\| \label{sdp_extended1.more.gen}\\ 
\textrm{subject to} \hspace{2mm} & 
\mathcal{Z}_0 Y_1 = \begin{bmatrix} P_1 \\ 0_{(S-n-m)\times (n+m)} \end{bmatrix} 
\label{sdp_extended2.more.gen}\\
& \begin{bmatrix} P_1 & (\Xi_1 Y_1)^\top  \\ 
\Xi_1 Y_1 & P_1 \end{bmatrix}  \succ 0 \label{eq:2SDP2.more.gen} \\
& \mathcal{Z}_0 G_2 = \begin{bmatrix} 0_{(n+m)\times (S-n-m)} \\ I_{S-n -m} 
\end{bmatrix} \,. \label{sdp_extended5.more.gen}
\end{alignat}
\end{subequations}
If this SDP is feasible then the dynamical controller 
\begin{equation}\label{extended.controller.more.gen}
\begin{array}{rl}
u^+= \left[ \begin{array}{c|c}
\overline{\mathcal{K}} & \hat{\mathcal{K}} \end{array} \right]
\left[ \begin{array}{cc} \xi \\ 
\hline \mathcal{Q}(\xi)
\end{array} 
\right] & \textrm{with} \\[0.3cm] 
\left[ \begin{array}{c|c}
\overline{\mathcal{K}} & \hat{\mathcal{K}} 
\end{array} \right] = V_0 \left[
\begin{array}{c|c} Y_1 P_1^{-1} & G_2 \end{array}
\right] &
\end{array}
\end{equation}
renders the origin of the closed-loop system an asymptotically stable equilibrium. \quad $\Box$
\end{corollary}
	
\emph{Proof.} The proof follows that of Theorem \ref{thm:approx}. 
The constraints \eqref{sdp_extended2.more.gen}, \eqref{sdp_extended5.more.gen}, 
along with $P_1\succ 0$ guaranteed by \eqref{eq:2SDP2.more.gen}, imply that $\mathcal{Z}_0 
\begin{bmatrix} G_1& G_2 \end{bmatrix}= I_{S}$, 
having set $G_1:=Y_1 P_1^{-1}$. 
Bearing in mind the expression of $\mathcal{K}:= \begin{bmatrix}\,\overline{\mathcal{K}} 
& \hat{\mathcal{K}} \end{bmatrix}$ in \eqref{extended.controller.more.gen}, we obtain 
\begin{equation}
\label{eq:GK.extended}
\begin{bmatrix}
V_0 \\ \mathcal{Z}_0 \end{bmatrix}
\begin{bmatrix}
G_1& G_2
\end{bmatrix}
= 
\begin{bmatrix}
\mathcal{K} \\ I_{S}
\end{bmatrix}\,.
\end{equation}
Finally, system \eqref{system.g(x).more.gen} with the control law 
\eqref{extended.controller.more.gen} can be written as 
$\xi^+= (\mathcal{A} +\mathcal{B} \mathcal{K}) \mathcal{Z}(\xi)$, or, in view of the identities 
\eqref{eq:GK.extended} and $\Xi_1 = \mathcal{A} \mathcal{Z}_0 + \mathcal{B} V_0$, as 
$\xi^+= \begin{bmatrix} \Xi_1 G_1& \Xi_1 G_2 \end{bmatrix} \mathcal{Z}(\xi)$.
The constraint \eqref{eq:2SDP2.more.gen} ensures that $\Xi_1 G_1$, the matrix describing 
the linear dynamics of the closed-loop system, is Schur,
and the thesis follows because by hypothesis $\mathcal{Q}(\xi)$ decays faster than linearly as $\xi$ goes to zero. 
\quad $\blacksquare$

As before, we can replace the property $\lim_{|\xi|\to 0} \frac{| \mathcal{Q}(\xi)|}{ |\xi |} =0$ 
by requiring $\mathcal{Q}(\xi)$ to be differentiable at $\xi=0$ and $\mathcal{Q}(0)=0$, so that 
$\mathcal{Q}(\xi)= \left[\frac{\partial \mathcal{Q}}{\partial \xi }\right]_{\xi =0} \xi + r(\xi)$,
with $r(\xi)$ differentiable and such that $\lim_{|\xi|\to 0} \frac{| r(\xi) |}{ |\xi |} =0$. 
In such a way, one can take
$\mathcal{Z}(\xi)=\big[\begin{smallmatrix} \xi \\ r(\xi) \end{smallmatrix}\big]$ instead of \eqref{Z(x,u)}. 
Further, the Lyapunov function $V(\xi)= \xi^\top P_1^{-1}\xi$ 
in Corollary \ref{cor:approx.extended.more.gen} 
can be used to estimate the ROA of the closed-loop system 
\eqref{system.g(x).more.gen}, \eqref{extended.controller.more.gen}, similarly to what 
has been done to establish Proposition \ref{prop:RoA}. 

\textbf{Example 5.} 
Consider the Euler discretization of an inverted pendulum
\begin{subequations} 
\label{eq:sys_example_pendulum_g(x)}
\begin{alignat}{2}
& x_1^+ = x_1 + T_s x_2 \\   
& x_2^+ = \displaystyle  \frac{T_s  g}{\ell} \sin x_1 + \left( 1 - \frac{T_s  \mu}{m \ell^2} \right) x_2 +
\frac{T_s }{m \ell}\cos x_1\, u \,,
\end{alignat}
\end{subequations}
where now the force is applied at the base, and this results in a state-dependent input vector field 
$\begin{bmatrix} 0 & \frac{T_s}{m \ell}\cos x_1\end{bmatrix}{}^\top$. The parameters 
$T_s, m, \ell, \mu, g$ and the states $x_1, x_2$ are the 
same as in Example 1. The problem is again that of stabilizing the unstable equilibrium in 
$(x, u)=(0,0)$.

\begin{figure}
\hspace*{-0.5cm}
{\includegraphics[width=7cm]{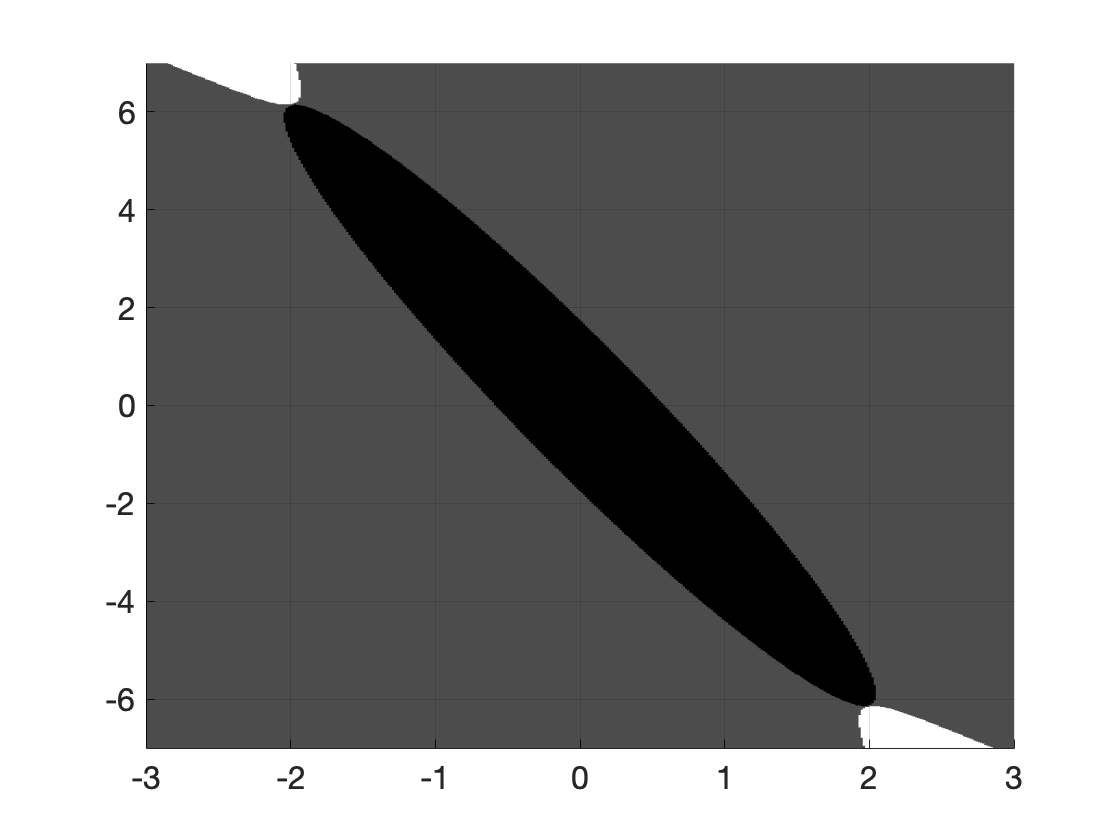}} 
\caption{
Results for Example 5. 
The grey set represents the set $\mathcal V$ where
$V(\xi^+)-V(\xi)$ is negative. Here, 
$\mathcal Z(\xi)=\left[\begin{smallmatrix} \xi^\top & \sin \xi_1-\xi_1 & 
(\cos \xi_1 -1)\, \xi_3\end{smallmatrix}\right]{}^\top$ and $V(\xi)=
\xi^\top P_1^{-1} \xi$, with $P_1^{-1}=\left[\begin{smallmatrix} 
    0.2159 &   0.0689 &   0.0123 \\
    0.0689 &   0.0240  &  0.0039 \\
    0.0123 &   0.0039  &  0.0009 \\
\end{smallmatrix}\right]$. 
The black set is a Lyapunov sublevel set $\mathcal R_{\gamma}$ 
contained in $\mathcal V$, hence it provides an estimate of the ROA for the system. 
and $\gamma=0.076$. Both sets $\mathcal V$ and $\mathcal R_{\gamma}$ 
are projected onto the plane $\{\xi: \xi_3=0\}$. 
}
\label{fig:ex5+}
\end{figure}

The vector $\mathcal{Q}(\xi)$ suggested by physical considerations is 
$\left[\begin{smallmatrix} \sin \xi_1 & \cos \xi_1 \, \xi_3\end{smallmatrix}\right]{}^\top$, 
which is zero at $\xi=0$ and differentiable. Hence, the function 
$r(\xi)=\left[\begin{smallmatrix} \sin \xi_1-\xi_1 & (\cos \xi_1 -1)\, \xi_3\end{smallmatrix}\right]{}^\top$ 
satisfies $\lim_{|\xi|\to 0} \frac{| r(\xi) |}{ |\xi |} =0$. 
Here, $r(\xi)$ 
is a preferred choice over  $\mathcal{Q}(\xi)$ because it yields a controllable linear part, which is 
necessary for the feasibility of the SDP.  We collect data by running an experiment with input uniformly 
distributed in $[-0.5,0.5]$, and with an initial state within the same interval. 
We collect $T=10$ samples corresponding to the motion of the pendulum that oscillates around 
the upright position. The SDP \eqref{eq:2SDP.extended.more.gen} is feasible and we obtain 
$\mathcal{K} = \begin{bmatrix} 
-17.6197 &  -5.6815 &  -0.3012 &   0 &  0 \end{bmatrix}$.
The controller locally asymptotically stabilizes 
the closed-loop system around the origin. 
For this controller, we numerically determine the set 
$\mathcal V = \{\xi: V(\xi^+)-V(\xi)=H(\xi) < 0\}$, with $H(\xi) := 
(\Xi_1 G_1 \xi +\Xi_1 G_2  \mathcal{Q}(\xi))^\top 
P_1^{-1} (\Xi_1 G_1 \xi +\Xi_1 G_2 \mathcal{Q}(\xi)) - \xi^\top P_1^{-1} \xi$, over
which the Lyapunov function $V(\xi)=\xi^\top P_1^{-1} \xi$ decreases. 
Any sub-level set $\mathcal R_\gamma$ of $V$
contained in $\mathcal V \cup \{0\}$ gives an estimate of the ROA for the closed-loop system. 
The set $\mathcal V$ and a sublevel set of $V$ are displayed in Figure \ref{fig:ex5+}. The 
$0$ values taken on by the last two entries of $\mathcal{K}$ 
(which correspond to the subvector $\hat{\mathcal{K}}$ in \eqref{extended.controller.more.gen}) 
is a byproduct of the minimization of $\|\Xi_1 G_2\|$, 
which in turn imposes a small value of $\|V_0 G_2\|$, in view of the addition of the integrator 
($V_0$ equals the last $m$ rows of $\Xi_1$, 
therefore $\Xi_1 G_2=\left[
\begin{smallmatrix}
X_1 G_2 \\ V_0 G_2
\end{smallmatrix}
\right]$). 
\quad $\blacksquare$

Corollary \ref{cor:approx.extended.more.gen} is a direct extension of 
Theorem \ref{thm:approx} and allows the designer to deal with a more general 
class of nonlinear systems, including systems with state-dependent input vector fields. 
Nevertheless, if it is known that the input vector field is state-independent, 
it is preferable to use the design proposed by Theorem \ref{thm:approx}, 
which might guarantee a global stabilization result by a static feedback in case 
the solution attains a zero cost, as formalized in Theorem \ref{thm:exact_lift}. 

\section{Robustness to disturbances and neglected nonlinearities}
\label{sec:noise}

In this section, we discuss robustness to disturbances and/or neglected nonlinearities.
Consider a system in the form
\begin{eqnarray} \label{system_noisy}
x^+ = AZ(x)+Bu+Ed
\end{eqnarray}
where $d \in \mathbb R^s$ is an unknown signal that accounts for process disturbances
and/or neglected nonlinearities (when $Z$
does \emph{not} include all the nonlinearities present in the system), 
whereas $E \in \mathbb R^{n \times s}$ 
is a known matrix that specifies which channel the signal $d$ enters. If such information
is not available then we simply let $E=I_n$. Because of $d$, the previous tools must be
modified to maintain stability guarantees. While the tools we use 
to study process disturbances and neglected nonlinearities are similar, we will
tackle the two cases separately. 

\subsection{Process disturbances: noisy data and robust invariance} \label{subsec:noise}

We start with the case where $d$ is a process disturbance.
The presence of $d$ affects the analysis in two different directions. First,
it affects controller design since it corrupts the data.\footnote{By following 
\cite[Section V-A]{de2019formulas}, the analysis can be extended to the case of measurement 
noise. We omit the details for brevity.} Second, it leads to notions other than Lyapunov stability
and ROA. We will address both the questions.
 
Similarly to the disturbance-free case, suppose we perform an experiment on the 
system, and we collect state and input samples satisfying $x(k+1)=AZ(x(k))+Bu(k)+Ed(k)$,
$k=0,\ldots,T-1$. These samples are then grouped into the data matrices $U_0, X_0, X_1, Z_0$ 
as in \eqref{eq:data}. Furthermore, let 
\begin{equation} \label{eq_D0}
D_0 := \begin{bmatrix} d(0) & d(1) & \cdots & d(T-1)  \end{bmatrix}  
\end{equation}
be the (unknown) data matrix that collects the samples of $d$. Our first step
is to establish an analogue of Lemma \ref{lem:main}.

\begin{lem} \label{lem:main2}
Consider any matrices $K \in \mathbb R^{m \times S}$,
$G \in \mathbb R^{T \times S}$ satisfying \eqref{eq:GK}.
Let $G$ be partitioned as $G = \begin{bmatrix} G_1 & G_2 \end{bmatrix}$, where
$G_1 \in \mathbb R^{T \times n}$. 
System \eqref{system_noisy} under the control law $u=KZ(x)$ results in
the closed-loop dynamics 
\begin{equation} \label{eq:GK_closed}
x^+ = \Psi x + \Xi Q(x) + Ed
\end{equation}
where $\Psi:= (X_1-ED_0) G_1$ and $\Xi:= (X_1-ED_0) G_2$,
\quad $\Box$
\end{lem}  

\emph{Proof.} 
Similarly to \eqref{eq:closed_loop_ideal}, we have
\begin{subequations}
\label{eq:closed_loop_ideal2}
\begin{alignat}{4}
x^+ &= \begin{bmatrix} B & A \end{bmatrix} \begin{bmatrix} K \\ I_S \end{bmatrix} Z(x) + Ed \\
&= \begin{bmatrix} B & A \end{bmatrix} \begin{bmatrix} U_0 \\ Z_0 \end{bmatrix} G Z(x) + Ed \\
&= (X_1-ED_0) G Z(x) + Ed \,.
\end{alignat}
\end{subequations}
The last identity follows as $X_1, U_0, Z_0, D_0$
satisfy the relation 
$x(k+1) = A Z(x(k)) + Bu(k)+E d(k)$, $k=0,\ldots,T-1$, which gives 
$X_1 = AZ_0+BU_0+E D_0$.
\quad $\blacksquare$ 

By looking at \eqref{eq:closed_loop_ideal2} we note that the closed-loop dynamics 
now depends on the unknown matrix $D_0$, and \eqref{eq:2SDP} 
no longer provides stability guarantees. In fact, the constraint \eqref{eq:2SDP2}
ensures that $M=X_1G_1$ is Schur. By Lemma \ref{lem:main2}, however,
the matrix of interest is now $\Psi=(X_1-ED_0)G_1$, and  
stability of $M$ does not ensure that also $\Psi$ is stable. To have stability,
we need to modify \eqref{eq:2SDP2} accounting for the uncertainty induced by $D_0$. 
A simple and effective way to achieve this is to ensure that $(X_1-ED)G_1$ is stable \emph{for all}
the matrices $D$ in a given set $\mathcal D$ to which $D_0$ is deemed to belong
(this approach can in fact be viewed as a \emph{robust control} approach).
We will consider the set 
\begin{equation} \label{eq:noise_model}
\mathcal D := \{ D \in \mathbb R^{s \times T}: 
DD^\top \preceq \Delta \Delta^\top \}
\end{equation} 
with $\Delta$ a design parameter,
and enforce, in place of \eqref{eq:2SDP2},
\begin{equation} \label{eq:2SDP_noisy} 
 Y_1^\top (X_1-ED)^\top P_1^{-1} (X_1-ED) Y_1 - P_1 + \Omega 
\prec 0  \quad \forall D \in \mathcal D
\end{equation}
where $Y_1$ and $P_1\succ0$ are decision variables which satisfy the identity $Y_1 P_1^{-1}=G_1$, 
while $\Omega\succ0$ is a free design parameter we will comment on shortly. 
By enforcing \eqref{eq:2SDP_noisy} we guarantee that 
$(X_1-ED)G_1$ is stable for all $D \in \mathcal D$, hence we ensure stability of 
$(X_1-ED_0)G_1$ if $D_0 \in \mathcal D$. 
The choice of the set $\mathcal D$ clearly reflects our prior information or guess about $d$.
For instance, if we know that $|d| \leq \delta$ for some $\delta > 0$
then we let $\Delta:=\delta \sqrt{T} I_s$. Stochastic disturbances can also be accounted for 
(possibly, with other choices of $\Delta$), see Section \ref{sec:noise_statistics}.
In general, large sets $\mathcal D$ make condition $D_0 \in \mathcal D$ easier to hold
but make \eqref{eq:2SDP_noisy} more difficult to satisfy. We proceed by 
making the assumption $D_0 \in \mathcal D$ explicit.

\begin{assumption} \label{ass:D0}
$D_0 \in \mathcal D$. \quad $\Box$
\end{assumption}

A final comment regards the matrix $\Omega$. This matrix ensures 
that $Y_1^\top (X_1-ED)^\top P_1^{-1} (X_1-ED) Y_1 - P_1$
is bounded away from singularity, as we vary $D$, by a \emph{known} quantity, and this is key to have 
an explicit expression for the ROA. There is no loss of 
generality in considering \eqref{eq:2SDP_noisy} instead of
\begin{equation} \label{eq:2SDP_noisy_weak} 
Y_1^\top (X_1-ED)^\top P_1^{-1} (X_1-ED) Y_1 - P_1 \prec 0 \quad \forall D \in \mathcal D. 
\end{equation}
Indeed, for any $\Omega \succ 0$ there exist 
$Y_1,P_1\succ 0$ that satisfy \eqref{eq:2SDP_noisy} if and only if there exist $Y_1,P_1\succ 0$
that satisfy \eqref{eq:2SDP_noisy_weak}. 

Condition \eqref{eq:2SDP_noisy} cannot be implemented directly as it involves 
\emph{infinitely} many constraints.
The next result provide a tractable (and convex) condition for \eqref{eq:2SDP_noisy}. 
Following \cite[Lemma A.4]{Petersen1986}\footnote{Lemma A.4 in \cite{Petersen1986},
also known as the \emph{Petersen's lemma}, permits to study matrix inequalities
which involve uncertainty, like \eqref{eq:2SDP_noisy}, 
and gives conditions under which such inequalities can be equivalently assessed 
considering only the `boundary' of the uncertainty, like \eqref{eq:Petersen} does.
We refer the reader to \cite{Bisoffi2021Petersen} for a recent discussion on the use of 
Petersen's lemma in data-driven control of linear and polynomial systems.},
we could actually establish the \emph{equivalence} between the next 
\eqref{eq:Petersen} and \eqref{eq:2SDP_noisy}. 
Here, we will only show that \eqref{eq:Petersen} implies \eqref{eq:2SDP_noisy},
which is enough for our purposes.  

\begin{lem} \label{lem:Petersen}
Suppose that there exist $Y_1 \in \mathbb R^{T \times n},
P_1 \in \mathbb S^{n \times n},$ and a scalar $\epsilon > 0$ such that 
\begin{eqnarray} \label{eq:Petersen} 
\left[ \begin{array}{ccc}
P_1-\Omega  & (X_1Y_1)^\top & Y_1^\top \\[0.1cm] 
X_1 Y_1 & P_1-\epsilon E\Delta \Delta^\top E^\top  & 0_{n \times T} \\[0.1cm]
Y_1 & 0_{T \times n} & \epsilon I_T
\end{array}
\right] \succ 0
\end{eqnarray}
\end{lem}
with $\Omega \succ 0$ and $\Delta$ given. Then,
\eqref{eq:2SDP_noisy} holds. \quad $\Box$ 

\emph{Proof.} See Appendix \ref{sec:petersen}. \quad $\blacksquare$ 

We arrive at the following main result. 

\begin{theorem} \label{thm:approx_noise}
Consider a nonlinear system as in \eqref{system_noisy} with $Z$
satisfying the condition \eqref{eq:limitQx} and with $d$ a process disturbance.
For a given $\Omega \succ 0$ and $\Delta$,
suppose that the following SDP (this is just 
\eqref{eq:2SDP} with \eqref{eq:2SDP2} replaced by \eqref{eq:Petersen}
to account for robust stability)
\begin{subequations}
\label{eq:2SDPnoisy}
\begin{alignat}{2}
\textrm{minimize}_{P_1,Y_1,G_2} \quad
& \|X_1G_2\| \\ 
\textrm{subject to} \quad 
& \eqref{eq:2SDP1}, \eqref{eq:Petersen}, \eqref{eq:2SDP4} 
\end{alignat}
\end{subequations} 
is feasible. If Assumption \ref{ass:D0} holds
then the control law $u=KZ(x)$ with $K$ in \eqref{eq:K_SDP} 
renders the origin an asymptotically stable equilibrium for the closed-loop system. \quad $\Box$
\end{theorem} 

\emph{Proof.} Suppose that \eqref{eq:2SDPnoisy} is feasible. 
Let $G_1=Y_1P_1^{-1}$ and note that the two constraints \eqref{eq:2SDP1} and \eqref{eq:2SDP4} together 
yield $Z_0 \begin{bmatrix} G_1 & G_2 \end{bmatrix} = I_S$.
This relation, combined with \eqref{eq:K_SDP}, gives \eqref{eq:GK}. 
In view of Lemma \ref{lem:main2}, the closed-loop dynamics satisfies $x^+=\Psi x+ \Xi Q(x)+Ed$, with
$\Psi=(X_1-ED_0)G_1$. Next, we prove that $\Psi$ is Schur. By Lemma \ref{lem:Petersen} and since 
$D_0 \in \mathcal D$ by hypothesis, \eqref{eq:2SDP_noisy} holds for $D=D_0$. We have in 
particular $P_1^{-1} Y_1^\top (X_1-ED_0)^\top P_1^{-1} (X_1-ED_0) Y_1P_1^{-1} - P_1^{-1} \prec 0$.
By recalling that $Y_1P_1^{-1}=G_1$, we conclude that $\Psi$
is Schur. The result follows from \eqref{eq:limitQx}. \quad $\blacksquare$

Building on Theorem \ref{thm:approx_noise} it is possible to characterize 
regions of attractions as well as robust invariant sets \cite{Blanchini1999}. 
We start with the ROA as a preliminary step for robust invariance. 
Consider the closed-loop dynamics $x^+=\Psi x+ \Xi Q(x)$ where we set $d\equiv0$ 
since we consider the ROA, and let $V(x):=x^\top P_1^{-1} x$. 
We have
\begin{eqnarray} \label{eq:Lyap_noise}
&& \hspace{-0.7cm}  V(x^+) - V(x) = \nonumber \\ 
 && \qquad  \hspace{-0.7cm} 
 \underbrace{(\Psi x+\Xi Q(x))^\top P_1^{-1} (\Psi x+\Xi Q(x)) - x^\top P_1^{-1} x}_{=: s(x)} \quad
\end{eqnarray}
with $\Psi=(X_1-ED_0)G_1$, $\Xi=(X_1-ED_0)G_2$. We cannot proceed as
in the disturbance-free case because $\Psi$ and $\Xi$ are unknown. Nonetheless, 
we can upper bound $s(x)$ with a quantity that is computable from data alone.
First, we tackle $x^\top \Phi x$ where $\Phi :=P_1^{-1}-\Psi^\top P_1^{-1} \Psi$.
By Theorem \ref{thm:approx_noise}, \eqref{eq:2SDP_noisy} 
holds for $D=D_0$, namely $P_1 \Phi P_1 - \Omega  \succ 0$.
Premultiplying this inequality left and right by $P_1^{-1}$ gives
$ \Phi - P_1^{-1}\Omega P_1^{-1} \succ 0$, and hence $x^\top \Phi x \ge x^\top \underline \Phi x$ 
for all $x$, where $\underline \Phi := P_1^{-1} \Omega P_1^{-1}$. 
Accordingly, we have
\begin{equation*} \label{eq:Lyap_noise1_2}
 V(x^+) - V(x) \leq -x^\top \underline \Phi x + (2 \Psi x +\Xi Q(x))^\top P_1^{-1} \Xi Q(x) .
\end{equation*}
Bearing in mind the expressions of $\Psi$ and $\Xi$, and the fact that 
$\|D_0\|_2 \leq \|\Delta\|_2$, we can write
\begin{eqnarray} \label{eq:Lyap_noise2}
&& \hspace{-0.7cm}  V(x^+) - V(x) \leq \nonumber \\ 
 && \qquad  \hspace{-0.7cm} 
 \underbrace{- x^\top \underline \Phi x + \ell_1(x) +\ell_2(x) + \ell_3(x) + \ell_4(x)}_{=: \ell(x)}  
\end{eqnarray}
having set
\begin{subequations} 
\begin{alignat}{2}
& \ell_1(x) := ( 2 X_1 G_1 x + X_1 G_2 Q(x) )^\top P_1^{-1} X_1 G_2 Q(x), \nonumber \\
& \ell_2(x) := \|\Delta\|_2 | ( 2 X_1 G_1 x + X_1 G_2 Q(x) )^\top P_1^{-1} E | | G_2 Q(x) |, \nonumber \\
& \ell_3(x) := \|\Delta\|_2 | 2 G_1 x + G_2 Q(x) | | E^\top P_1^{-1} X_1 G_2 Q(x) |, \nonumber \\
& \ell_4(x) := \|\Delta\|_2^2 \|E^\top P_1^{-1} E\|_2 | 2 G_1 x + G_2 Q(x) | | G_2 Q(x) |, \nonumber 
\end{alignat}
\end{subequations}
which are all computable from data alone.


\begin{proposition} \label{prop:ROA_noise}
Consider the same setting as in Theorem \ref{thm:approx_noise}.
Let $\mathcal L := \{x: \ell(x) < 0\}$, with $\ell(x)$ as in \eqref{eq:Lyap_noise2},
and consider the Lyapunov function $V(x)=x^\top P_1^{-1} x$. Then, 
any sub-level set $\mathcal R_\gamma := \{x: V(x) \leq \gamma \}$ of $V$ 
contained in $\mathcal L \cup \{0\}$ is a PI set for the closed-loop system with $d \equiv 0$
and defines an estimate of the ROA relative to $\overline x = 0$.
\quad $\Box$
\end{proposition}

We now consider robust invariance \cite[Definition 2.2]{Blanchini1999}.

\begin{definition} \label{def:RPI}
A set $\mathcal S$ is called \emph{robustly positively invariant} (RPI) for the system $x^+ = f(x,d)$
if for every $x(0) \in \mathcal S$ and all $d(t) \in \mathcal I$, with $\mathcal I$ a compact set,
the solution is such that $x(t) \in \mathcal S$ for $t>0$. \quad $\Box$
\end{definition}

Unlike local stability and invariance, 
which pose conditions on the disturbance only relatively to the data collection phase
(Assumption \ref{ass:D0}, i.e. the condition $D_0 \in \mathcal D$), 
robust invariance constrains $d$ for \emph{all} times $t\geq0$.
This calls for strengthening Assumption \ref{ass:D0}
 in the sense of Definition \ref{def:RPI}.\,\footnote{As an example, 
a Gaussian disturbance may satisfy the condition $D_0 \in \mathcal D$ but is not 
bounded in the sense of Definition \ref{def:RPI}. Set invariance for
unbounded disturbances is studied in \cite{Kofman2012}. 
We will not pursue this problem here.
}

\begin{assumption} \label{ass:RPI}
$|d| \leq \delta$ for some known $\delta>0$. \quad $\Box$
\end{assumption}  

Assumption \ref{ass:RPI} is indeed stronger than Assumption \ref{ass:D0} in the
sense that it implies Assumption \ref{ass:D0} once we set $\Delta := \delta \sqrt{T} I_s$.
We can now proceed with the analysis of robust invariance. 
Consider the closed-loop system $x^+=\Psi x+ \Xi Q(x) + E d$ with 
$d$ satisfying Assumption \ref{ass:RPI}, and let $V(x):=x^\top P_1^{-1} x$. It is simple to verify 
that we now have
{\setlength\arraycolsep{2.0pt}
\begin{eqnarray} \label{eq:Lyap_noise_RPI} 
V(x^+) - V(x) \leq \ell(x) + g(x,\delta) ,
\end{eqnarray}}%
where $\ell(x)$ is as in \eqref{eq:Lyap_noise2}, and where
\begin{subequations} 
\label{eq:gxdelta}
\begin{alignat}{2} 
& g(x,\delta) := r_1(x) \delta + r_2(x) \delta + r_3 \delta^2 , \\
& r_1(x):=2 | (X_1 G_1 x + X_1 G_2 Q(x))^\top P_1^{-1} E | , \\
& r_2(x):=2 \|\Delta\|_2 \| E^\top P_1^{-1} E \|_2 | G_1 x + G_2 Q(x) |, \\
& r_3:=\| E^\top P_1^{-1} E \|_2 .  
\end{alignat}  
\end{subequations} 
Let 
\begin{equation} 
\mathcal X := \{ x: \ell(x) + g(x,\delta) \leq 0  \} \label{eq:RPI_set_X} 
\end{equation} 
and let $\mathcal X^c$ be its complement. 

\begin{theorem} \label{thm:RPI} 
Consider a nonlinear system as in \eqref{system_noisy} with $Z$
satisfying \eqref{eq:limitQx} and with $d$ a process disturbance
for which Assumption \ref{ass:RPI} holds. 
For a given $\Omega \succ 0$,
suppose that \eqref{eq:2SDPnoisy}
is feasible with $\Delta := \delta \sqrt{T} I_s$, and consider the control law $u=KZ(x)$ 
where $K$ is as in \eqref{eq:K_SDP}.
Let $V(x):=x^\top P_1^{-1} x$, and define $\mathcal R_\gamma := \{ x: V(x) \leq \gamma \}$, 
where $\gamma>0$ is arbitrary. Finally, let
$\mathcal Z :=  \mathcal R_\gamma \cap \mathcal X^c$ 
($\mathcal Z$ defines all the points $x$ of $\mathcal R_\gamma$ for which 
the Lyapunov difference $V(x^+)- V(x)$ can be positive;
it is nonempty for any choice of $\gamma>0$).
If
\begin{equation} \label{eq:main_RPI}
V(x) + \ell(x) + g(x,\delta) \leq \gamma \quad \forall x \in {\mathcal Z}
\end{equation} 
then $\mathcal R_\gamma$ is an RPI set for the closed-loop system.
\quad $\Box$
\end{theorem} 

\emph{Proof.} As shown in Theorem \ref{thm:approx_noise}, 
feasibility of \eqref{eq:2SDPnoisy}, along with $D_0 \in \mathcal D$, 
ensures that $V(x)=x^\top P_1^{-1} x$ is a Lyapunov function for the 
linear part of the dynamics, and \eqref{eq:limitQx} ensures
that $\mathcal L = \{x: \ell(x) <0\}$, with $\ell(x)$ as in \eqref{eq:Lyap_noise2}, is nonempty 
(if $\mathcal L$ is empty then \eqref{eq:main_RPI} never holds).
Then, assume that \eqref{eq:main_RPI} holds
and let $x \in \mathcal R_\gamma$. We divide the analysis in two cases.
First assume that $x \notin \mathcal Z$. Since $x \in \mathcal R_\gamma$ then
$x \notin \mathcal X^c$. Then $x \in \mathcal X$, so that 
$V(x^+)-V(x) \leq \ell(x) + g(x,\delta) \leq  0$, and this implies $x^+ \in \mathcal R_\gamma$.
Next, assume that $x \in \mathcal Z$. In view of \eqref{eq:main_RPI} we
have $V(x^+) \leq \gamma$, 
thus $x^+ \in \mathcal R_\gamma$.
 \quad $\blacksquare$

Equations \eqref{eq:Lyap_noise2} and \eqref{eq:Lyap_noise_RPI} suggest that from a practical point of view it
might be convenient to \emph{regularize} the objective function in \eqref{eq:2SDPnoisy}
so as to mitigate the effect of the disturbance. As shown in the subsequent 
numerical examples, a convenient  choice is the following one:
\begin{subequations}
\label{eq:4SDP}
\begin{alignat}{2}
\textrm{minimize}_{P_1,Y_1,G_2} \quad
& \|X_1 G_2\| + \lambda_1 \|P_1\| + \lambda_2 \|G_2\|  \\ 
\textrm{subject to} \quad 
& \eqref{eq:2SDP1}, \eqref{eq:Petersen}, \eqref{eq:2SDP4} 
\end{alignat}
\end{subequations} 
where $\lambda_1,\lambda_2 \geq 0$ are weighting parameters. 
Penalizing $\|P_1\|$ increases the smallest eigenvalue of $\underline \Phi$,
while penalizing $\|G_2\|$ decreases the various terms $\ell_i$ and $r_i$ in \eqref{eq:Lyap_noise2}
and \eqref{eq:Lyap_noise_RPI}.
Notice that penalizing $\|P_1\|$ might increase the terms $\ell_i$ and $r_i$, but while 
these quantities depend on $P_1^{-1}$, $\underline \Phi$ depends on $P_1^{-2}$,
so penalizing $\|P_1\|$ can still be advantageous. 
 
Since \eqref{eq:4SDP} has the same feasible set as \eqref{eq:2SDPnoisy}
it is understood that all the results of this section as well as those to follow 
remain true if \eqref{eq:2SDPnoisy} is replaced with \eqref{eq:4SDP}.

\begin{figure}[h!]
\normalsize
\caption{
Simulation results for \eqref{eq:4SDP} with
$Z(x)=\begin{bmatrix} x_1 \,\,\, x_2 \,\,\, \sin(x_1) - x_1 \end{bmatrix}{}^\top$,
$\lambda_1=\lambda_2=0.1$, and $\delta=0.01$.
Left: the grey set represents the set $\mathcal X$ in \eqref{eq:RPI_set_X},
while the blue set is the RPI set $\mathcal R_\gamma$; here, 
$P_1^{-1}=[\begin{smallmatrix} 0.1901 & 0.0664 \\ 0.0664 & 0.0475 \end{smallmatrix}]$
and $\gamma=0.4440$. 
The black set wrapping $\mathcal R_\gamma$ is
the ROA, which is larger than the RPI set. Finally, the red set around 
the origin corresponds to the set $\mathcal Z$; here,
$\max_{x \in \mathcal Z} \, V(x) + \ell(x) + g(x,\delta) =0.001$.
States originating in $\mathcal Z$ do not exit $\mathcal R_\gamma$.
In particular, any sub-level set $\mathcal R_\gamma = \{ x: V(x) \leq \gamma \}$ 
with $\gamma \in [0.0010,0.4440]$ is an RPI set for the closed-loop system.
Right: zoom showing $\mathcal R_\gamma$ close to the border 
of $\mathcal X$.
}
\label{fig:pendulum_RPI} 
\hspace*{-0.5cm}
{\includegraphics[width=4.8cm]{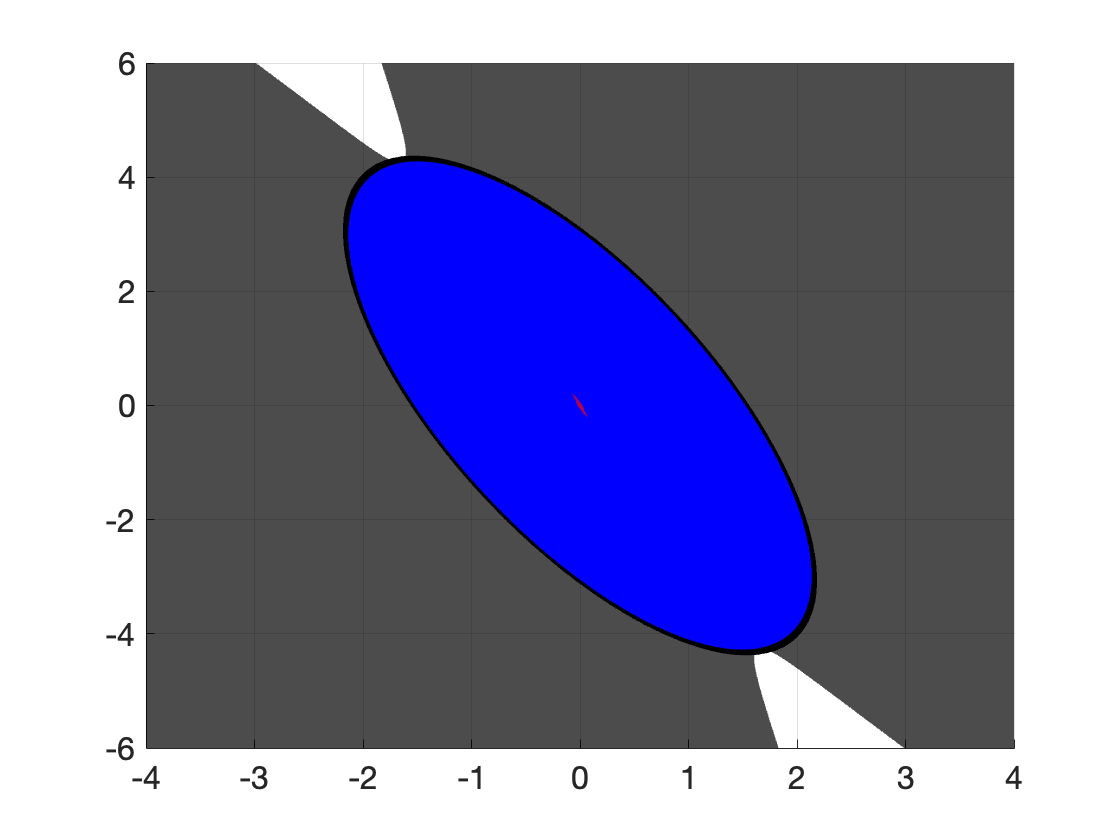}} \hspace*{-0.6cm}
{\includegraphics[width=4.8cm]{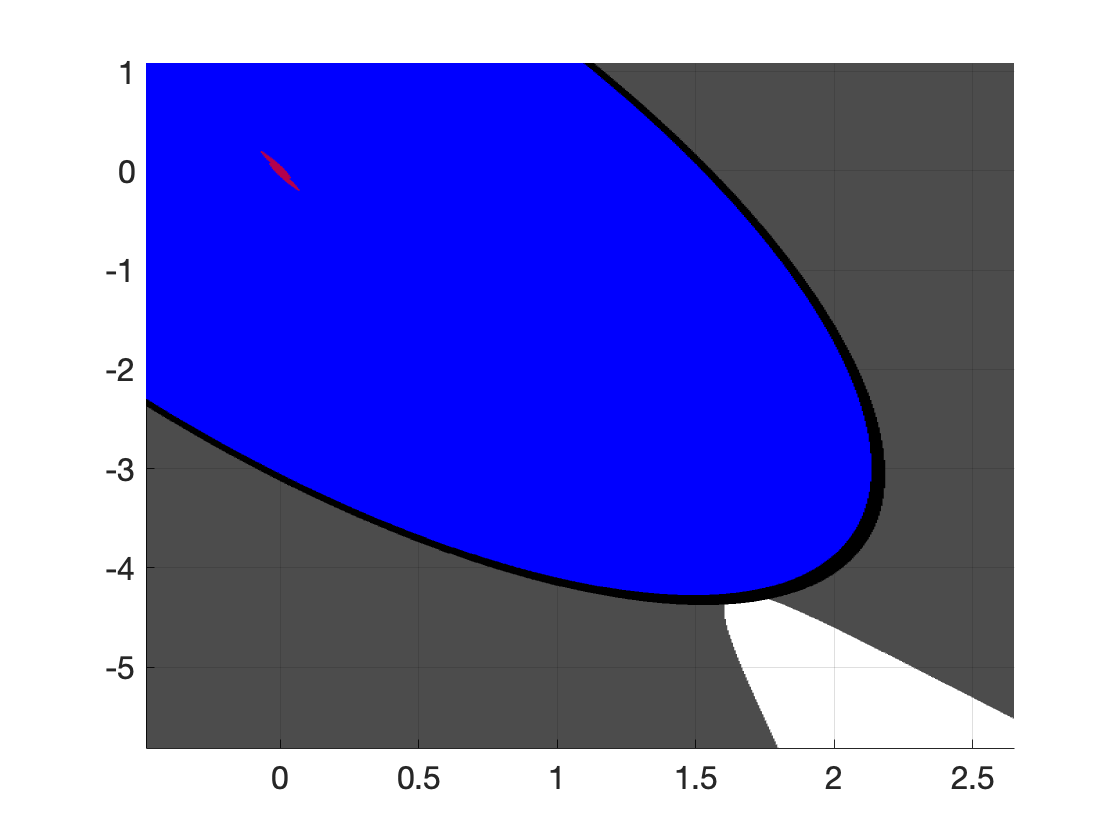}} 
\end{figure}

\textbf{Example 6.} We consider again the inverted pendulum of 
Example 1, this time assuming that a disturbance $d$ acts on the control
channel, namely we have $E = [\begin{smallmatrix} 0 \\ 1 \end{smallmatrix}]$ and
the second equation is modified as
\begin{equation*}
x_2^+ = \displaystyle  \frac{T_s  g}{\ell} \sin x_1 + 
\left( 1 - \frac{T_s  \mu}{m \ell^2} \right) x_2 +
\frac{T_s}{m \ell^2}u + d .
\end{equation*}
We collect data by running
an experiment with input uniformly distributed in $[-0.5, 0.5]$,
and with an initial state within the same interval. 
We consider a disturbance uniformly distributed in $[-\delta, \delta]$.
We collect $T=30$ samples and solve \eqref{eq:4SDP} with 
$\lambda_1=\lambda_2=0.1$, $\Omega = I_2$ and 
$\Delta =\delta \sqrt{T}$. Figure \ref{fig:pendulum_RPI} reports
results for $\delta=0.01$. We observe the following:
(i) \eqref{eq:4SDP} remains feasible up to $\delta \approx 0.1$ but for such large 
values we get empty estimates of ROA/RPI. (ii) the regularization is in fact
needed to get nonempty estimates of ROA/RPI, and even small values for $\lambda_1,\lambda_2$
suffice. This permits to preserve the baseline strategy of nonlinearity minimization.
In fact, the controller we obtain is
$K=\begin{bmatrix} -23.9436 & -11.4581 &  -9.8564 \end{bmatrix}$,
which generates the term $-9.8564 \sin(x_1)$ that 
approximately cancels out the nonlinearity. (iii) Compared with the disturbance-free case,
here we need a larger number of samples to get nonempty estimates of ROA/RPI,
although \eqref{eq:4SDP} remains feasible even for $T=10$. 
Intuitively,
collecting more samples can indeed help to get more information on the system's dynamics;
we will elaborate on this point in Section \ref{sec:noise_statistics}. 
\quad  $\blacksquare$

\subsection{Neglected nonlinearities} \label{subsec:neglected}

A similar analysis can be carried out in case of neglected nonlinearities.
The difference is that now $d$ will be a function of the state $x$, say $d=d(x)$. 
The \emph{combination} of neglected nonlinearities and genuine disturbances is 
also possible, but we omit the details for brevity. Thus, the analysis
which follows only considers {invariance} instead of \emph{robust} invariance.

In order to handle the case of neglected nonlinearities, we assume 
some knowledge on the \emph{strength} of such nonlinearities
(Assumption \ref{ass:setQ} is essentially the counterpart of Assumption \ref{ass:RPI}).

\begin{assumption} \label{ass:setQ}
We know a set $\mathcal Q \subseteq \mathbb R^n$ and a scalar $\delta>0$ such that 
$|d(x)| \leq \delta$ for all $x \in \mathcal Q$. \quad $\Box$
\end{assumption} 

\begin{theorem} \label{thm:PI_neglected}
Consider a nonlinear system as in \eqref{system_noisy} with $Z$
satisfying \eqref{eq:limitQx} and with $d=d(x)$ a nonlinear function of the state 
for which Assumption \ref{ass:setQ} holds. 
Consider an experiment on the system such that 
$x(k) \in \mathcal Q$ for $k=0,\ldots,T-1$. For a given $\Omega \succ 0$,
suppose that \eqref{eq:2SDPnoisy} is feasible with
$\Delta=\delta \sqrt{T} I_s$. Let $V(x):=x^\top P_1^{-1} x$ and 
$\mathcal R_\gamma := \{x: V(x) \leq \gamma \}$
where $\gamma>0$ is arbitrary. Finally,
let $\mathcal X$ be as in \eqref{eq:RPI_set_X}
and ${\mathcal Z} :=  \mathcal R_\gamma \cap {\mathcal X}{}^c$.
If $\mathcal R_\gamma \subseteq \mathcal Q$ and
\begin{equation} \label{eq:main_RPI_neglected}
V(x) + \ell(x) + g(x,\delta) \leq \gamma \quad \forall x \in {\mathcal Z}
\end{equation} 
then $\mathcal R_\gamma$ is a PI set for the closed-loop system.
\quad $\Box$
\end{theorem} 

\emph{Proof.} Under the stated conditions we have $D_0 \in \mathcal D$.
Thus, the feasibility of \eqref{eq:2SDPnoisy}
guarantees that $V(x)=x^\top P_1^{-1} x$ is a Lyapunov function for the 
linear part of the dynamics, and \eqref{eq:limitQx} ensures
that $\mathcal L = \{x: \ell(x) <0\}$, with $\ell(x)$ as in \eqref{eq:Lyap_noise2}, is nonempty (otherwise
\eqref{eq:main_RPI_neglected} would never hold).
Then, assume that \eqref{eq:main_RPI_neglected} holds and
let $x \in \mathcal R_\gamma$. Since $x \in \mathcal R_\gamma$ 
then $x \in \mathcal Q$, and therefore $|d(x)| \leq \delta$. Hence, exactly 
as in \eqref{eq:Lyap_noise_RPI}, we have
$V(x^+)-V(x) \leq \ell(x) + g(x,\delta)$ where $g(x,\delta)$ is as in \eqref{eq:gxdelta}.   
The rest of the proof is analogous to that of Theorem \ref{thm:RPI}. 
Assume that $x \notin \mathcal Z$. Since $x \in \mathcal R_\gamma$ then
$x \notin \mathcal X^c$. Thus $x \in \mathcal X$, and hence 
$V(x^+)-V(x) \leq \ell(x) + g(x,\delta) \leq  0$,
which implies $x^+ \in \mathcal R_\gamma$.
Next, assume that $x \in \mathcal Z$. In view of \eqref{eq:main_RPI_neglected}, we
have $V(x^+) \leq \gamma$, thus $x^+ \in \mathcal R_\gamma$.
\quad $\blacksquare$

We can also have asymptotic stability under a strengthened 
Assumption \ref{ass:setQ}. Here we report a prototypical result.

\begin{theorem} \label{thm:LAS_neglected}
Consider the same setting as in Theorem \ref{thm:PI_neglected},
and suppose that $|d(x)|\leq \delta(x)$ for all $x$,
where $\delta(x): \mathbb R^n \rightarrow \mathbb R_+$ 
is some known function such that $\lim_{|x|\to 0} \frac{\delta(x)}{|x|}=0$. 
Let $\ell(x)$ be as in \eqref{eq:Lyap_noise2},
and let $g(x,\delta(x))$ be as in \eqref{eq:gxdelta} with $\delta$ replaced by $\delta(x)$.
Finally, define $\mathcal W := \{x: \ell(x)+g(x,\delta(x))<0\}$.
Then, the origin is an asymptotically stable equilibrium for the closed-loop system,
and any set $\mathcal R_\gamma := \{ x: V(x) \leq \gamma \}$ 
of $V$ contained in $\mathcal W \cup \{0\}$ 
is a PI set and defines an estimate of the ROA relative to $\overline x=0$. \quad $\Box$
\end{theorem} 

\emph{Proof.} Analogously to \eqref{eq:Lyap_noise_RPI}, 
the Lyapunov function satisfies $V(x^+)-V(x) \leq \ell(x)+g(x,\delta(x))$ for all $x$. 
Then the result follows immediately. \quad $\blacksquare$

\textbf{Example 7.} Consider the previous example, but this 
time assume that we purposely neglect the nonlinearity and 
design a \emph{linear} control law. Specifically, the dynamics
of the inverted pendulum can be written as  
\begin{subequations} 
\label{eq:sys_example_pendulum_neglected}
\begin{alignat}{2}
x_1^+ & = x_1 + T_s x_2, \nonumber \\   
x_2^+ & = \displaystyle  \frac{T_s  g}{\ell} x_1 + 
\left( 1 - \frac{T_s  \mu}{m \ell^2} \right) x_2 +
\frac{T_s}{m \ell^2}u + d, \nonumber \\
d & =\frac{T_s  g}{\ell} (\sin x_1 - x_1). \nonumber 
\end{alignat}
\end{subequations}
In this case, the type of dynamics is known, hence we focus on  
Theorem \ref{thm:LAS_neglected}. We consider $\delta(x)=2 |\sin x_1 - x_1|$,
thus $|d(x)| \leq \delta(x)$ for all $x$ (we over-approximate $d$ by more than $100\%$).
We run an experiment with input and initial state
uniformly distributed in $[-0.1, 0.1]$. This ensures that up to $T=10$ the state $x_1$
remains close to the equilibrium, so that $d$ remains small. 
In particular, with this choice, $x_1$ never exceeds $\pm 0.06$
($\approx \pm 3.5^\circ$), and $\delta(x) \leq 3\cdot10^{-5}=:c$. Thus we take $T=10$,
set $\Omega = I_2$, $\Delta=c\sqrt{T}$ and solve \eqref{eq:4SDP} 
(by the same arguments in Example 6 
on the impact of noise on the estimate of the ROA/RPI, 
we solve the regularized version of \eqref{eq:2SDPnoisy}). 

\begin{figure}[t]
\normalsize
\caption{
Simulation results when we consider a linear control law. The grey set represents the set ${\mathcal W}$, 
while the black set represents the set $\mathcal R_\gamma$ which defines the ROA; here, 
$P_1^{-1}=[\begin{smallmatrix} 0.2116 & 0.1291 \\ 0.1291 & 0.1351 \end{smallmatrix}]$
and $\gamma=0.0473$.}
\label{fig:pendulum_RPI_neglected} 
\hspace*{-0.5cm}
{\includegraphics[width=7cm]{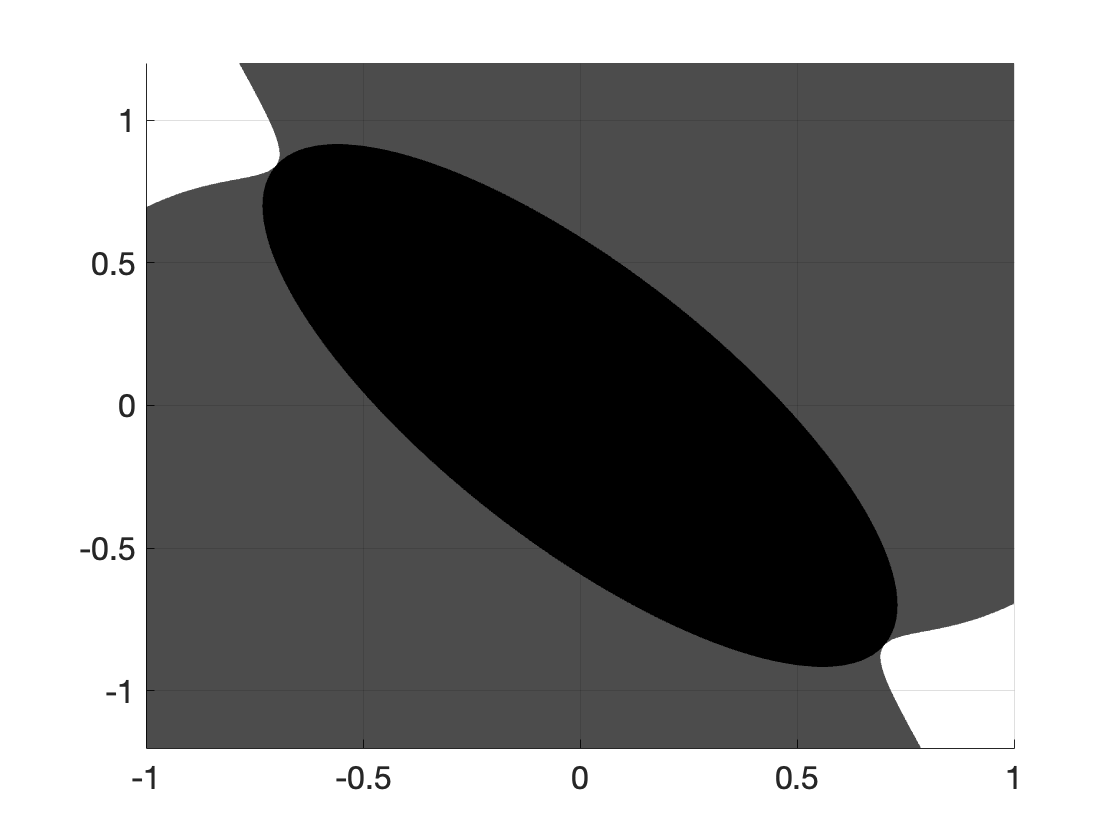}} 
\end{figure}

Note that \eqref{eq:4SDP} now involves only the variables $P_1,Y_1$, thus
only the two constraints \eqref{eq:2SDP1} and \eqref{eq:Petersen} 
are present. We get $K=\begin{bmatrix} -19.0204 & -10.7947 \end{bmatrix}$
and the ROA in Figure \ref{fig:pendulum_RPI_neglected}.
As expected, the outcome is worse than the one 
obtained when we exploit the knowledge of the nonlinearities
and we use a nonlinear control law. In particular, the main shortcoming is 
that we now need to run the experiment close to the equilibrium
in order to keep $d$ small, which is not needed when we take the nonlinearity 
into account.  \quad $\blacksquare$

\subsection{Results in probability} \label{sec:noise_statistics}

All previous results rest on the assumption that $D_0 \in \mathcal D$.
Clearly, once the experiment is performed and the data are collected,
whether $D_0 \in \mathcal D$  or not is a \emph{deterministic} property
(\emph{yes} or \emph{no}). Yet, certifying that $D_0$ actually belongs to $\mathcal D$ can be a 
difficult task. It turns out that we can establish results that 
relate closed-loop stability with the \emph{probability} that
$D_0 \in \mathcal D$. We focus on the case of process 
disturbances, in particular we give a probabilistic version of Theorem \ref{thm:approx_noise}.
 
\begin{theorem} \label{thm:approx_noise_prob}
Consider a nonlinear system as in \eqref{system_noisy} with $Z$
satisfying \eqref{eq:limitQx} and with $d$ a process disturbance.
For a given $\Omega \succ 0$ and $\Delta$, 
suppose that \eqref{eq:2SDPnoisy} is feasible.
If $D_0 \in \mathcal D$ with probability at least $p$ then the control law $u=KZ(x)$, with 
$K$ as in \eqref{eq:K_SDP}, renders the origin an asymptotically stable equilibrium
with probability at least $p$.
\quad $\Box$
\end{theorem} 

\emph{Proof.} The result is a direct consequence of the 
\emph{law of total probability} \cite[Theorem 3, pp. 28]{Lindgren1993}. Given two events 
$\mathcal E_1$ and $\mathcal E_2$, 
let ${P}(\mathcal E_1)$ and $P(\mathcal E_1|\mathcal E_2)$ 
denote the probability of $\mathcal E_1$ and the 
conditional probability of $\mathcal E_1$ given $\mathcal E_2$. Let $\mathcal E_1$ denote the event
that $K$ is stabilizing and $\mathcal E_2$ denote the event $D_0 \in \mathcal D$.
We have $P(\mathcal E_1) = P(\mathcal E_1|\mathcal E_2) P(\mathcal E_2)
+P(\mathcal E_1|\mathcal E_2^c) P(\mathcal E_2^c)$, with
$\mathcal E^c$ the complement of $\mathcal E$. Then, $P(\mathcal E_1) \geq 
P(\mathcal E_1|\mathcal E_2) P(\mathcal E_2)$
and the result follows because $P(\mathcal E_1|\mathcal E_2)=1$ by Theorem \ref{thm:approx_noise}.
\quad $\blacksquare$ 

Theorem \ref{thm:approx_noise_prob} allows us to extend our range of application 
to cases where bounds on $d$ are known only with a limited accuracy,
as exemplified in next Proposition \ref{prop:average}.
Theorem \ref{thm:approx_noise_prob} has another interesting implication. 
For disturbances obeying the
\emph{law of large numbers} \cite[Section 5]{Lindgren1993}
we can repeat the same experiment multiple times 
and average the data so as to filter out noise. 
Specifically, suppose we make $N$ experiments on system \eqref{system_noisy}, 
each of length $T$,
and let $( U_{0}^{(r)},D_{0}^{(r)},Z_{0}^{(r)},X_{1}^{(r)})$, with $r=1,\ldots,N$,
be the dataset resulting from the $r$-th experiment.
Given $N$ matrices $S^{(r)}$, with $r=1,\ldots,N$, let 
$\underline S := \frac{1}{N} \sum_{r=1}^N S^{(r)}$
denote their average. Since each dataset satisfies the
relation $X_1^{(r)} = A  Z_0^{(r)} + B U_0^{(r)} + E D_0^{(r)}$, if we average $N$ datasets
we obtain the relation
\begin{eqnarray} \label{eq:recursive_ave}
\underline X_1 = A  \underline Z_0 + B  \underline U_0 + E \underline D_0 
\end{eqnarray} 
Because the dynamics are nonlinear, \eqref{eq:recursive_ave} does \emph{not} represent a valid
trajectory of the system in the sense that it cannot result
from a single experiment on \eqref{system_noisy}. Yet, and this is the crucial point, the dataset
$(\underline U_0,\underline D_0,\underline Z_0,\underline X_1)$ still provides a data-based 
parametrization of the closed loop in the sense of Lemma \ref{lem:main2}. Specifically, for any
$K,G$ satisfying 
\begin{equation} \label{eq:GK_average}
\begin{bmatrix} K \\ I_S \end{bmatrix} = \begin{bmatrix} 
\underline U_0 \\ \underline Z_0 \end{bmatrix} G
\end{equation}
we have (\emph{cf.} \eqref{eq:closed_loop_ideal}) 
\begin{equation}
\label{eq:closed_loop_ideal_average}
A+BK = (\underline X_1 - E \underline D_0) G  .
\end{equation}
Hence, Lemma \ref{lem:main2}, and consequently 
Theorems \ref{thm:approx_noise} and
\ref{thm:approx_noise_prob}, apply to $(\underline U_0,
\underline D_0,\underline Z_0,\underline X_1)$ with
no modifications, with the advantage that $\underline D_0$ will have a 
reduced norm in expectation thanks to the {law of large numbers}.

While the law of large numbers gives an asymptotic result,
there are recent results in non-asymptotic statistics that permit
us, for relevant classes of disturbance, to get high-confidence bounds on 
$\|\underline D_0\|_2$ even
with a \emph{finite} number of experiments. As an example, we give the 
following result.\footnote{The notation
used in the sequel is standard, \emph{e.g.}, see \cite{Lindgren1993}. Independent 
and identically distributed random vectors are abbreviated as i.i.d..
We will denote by $\mathcal {N} (\mu,\Sigma)$ the multivariate 
normal (Gaussian) distribution with mean $\mu$ and covariance matrix $\Sigma$.}

\begin{proposition} \label{prop:average}
Consider $N$ experiments, each of length $T$, on system \eqref{system_noisy}, 
and assume that the disturbances $d(k) \in \mathbb R^s$ are i.i.d. zero-mean 
random vectors with covariance matrix $\Sigma$
such that $|d(k)| \leq \delta$ \emph{almost surely} (i.e., with probability $1$). 
Then, for all $\mu>0$, 
\begin{equation} \label{eq:Bernstein}
\|\underline D_0\|_2 \leq \sqrt{T \left( \frac{\|\Sigma\|_2}{N} + \mu \right)}
\end{equation}
with probability at least 
$1-2s\, \text{exp}\left( -\frac{T N \mu^2}{2\delta^2 (\|\Sigma\|_2 + N\mu)} \right)$.

Let instead the disturbances $d(k)$ be i.i.d. random vectors drawn from 
$\mathcal {N} (0,\Sigma)$. Then, for all $\mu>0$, 
\begin{equation} \label{eq:Wainwright}
\|\underline D_0\|_2 \leq \sqrt{\frac{T}{N}}  \left( \lambda_{\text{max}} (\Sigma^{1/2}) (1+ \mu  ) 
+ \sqrt{\frac{\text{trace}(\Sigma)}{T}} \right)
\end{equation}
with probability at least $1-\text{exp}( -T \mu^2/2)$.
where $\lambda_{\text{max}}$ denotes the
maximum eigenvalue. \quad $\Box$
\end{proposition}

\emph{Proof.} Since the disturbances $d(k)$ are independent then the vectors which 
form the columns of $\underline D_0$
are also independent. This can be easily verified, for instance, through the so-called
\emph{characteristic function}, \emph{e.g.}, see  \cite[Theorem 28, pp. 131]{Lindgren1993}.
It is also easy to verify that these vectors have zero mean and covariance matrix 
$\Sigma/N$. The bounds \eqref{eq:Bernstein} and \eqref{eq:Wainwright} follow from 
Corollary 6.20 and Theorem 6.1 in \cite{Wainwright2019}, respectively. 
\quad $\blacksquare$ 

Under the assumption on the disturbances stated 
in Proposition \ref{prop:average}, 
we can choose $\Delta=\eta I_s$ with $\eta$
equal to the right-hand side of \eqref{eq:Bernstein} or \eqref{eq:Wainwright},
and control $\eta$ via $T, \mu$ and $N$. 
This may lead us to satisfy, with a certain probability, the condition
$\|\underline D_0\|_2 \leq \eta$ (thus $\underline D_0 \in \mathcal D$) 
with $\eta$ small. As a result, we may render \eqref{eq:2SDPnoisy} easier to satisfy and 
have stability guarantees (in probability). Specifically, by applying
Theorem \ref{thm:approx_noise_prob}, if \eqref{eq:2SDPnoisy}, 
with $X_1, Z_0$ replaced by $\underline X_1, \underline Z_0$,
is feasible then the control law $u=KZ(x)$, where  
$K$ is given by \eqref{eq:K_SDP} with $U_0$ replaced by 
$\underline U_0$, will asymptotically stabilize the origin with the same probability 
as condition $\|\underline D_0\|_2 \leq \eta$ is satisfied.

A second advantage of having $\|\underline D_0\|_2 \leq \eta$ with $\eta$ small
is that, by virtue of \eqref{eq:Lyap_noise2} and \eqref{eq:Lyap_noise_RPI}, 
we may have (in probability) less conservative estimates for the ROA and RPI sets
compared to the ones obtained with deterministic (worst-case) bounds
for the disturbance. 

\textbf{Example 8.} We consider again Example 6 under the same 
experimental setup for the disturbance, but now we repeat the 
experiment $N=100$ times, each time
using the same input pattern. For the uniform distribution it holds that $\Sigma=\delta^2/3$. 
With $\mu=4\cdot10^{-5}$, Proposition \ref{prop:average} implies
$\|\underline D_0\|_2 \leq 0.0348$ with probability at least $99.48\%$.
The bound is much tighter compared to the worst-case bound
$\|\underline D_0\|_2 \leq  \delta \sqrt{T}=0.0548$ obtained by only exploiting 
the property $|d| \leq \delta$.
  
\begin{figure}[h!]
\normalsize
\caption{Simulation results for the pendulum in case of repeated 
experiments. See the caption of Figure \ref{fig:pendulum_RPI} for a description  
of the various sets.}
\label{fig:pendulum_RPI_average} 
{\includegraphics[width=7cm]{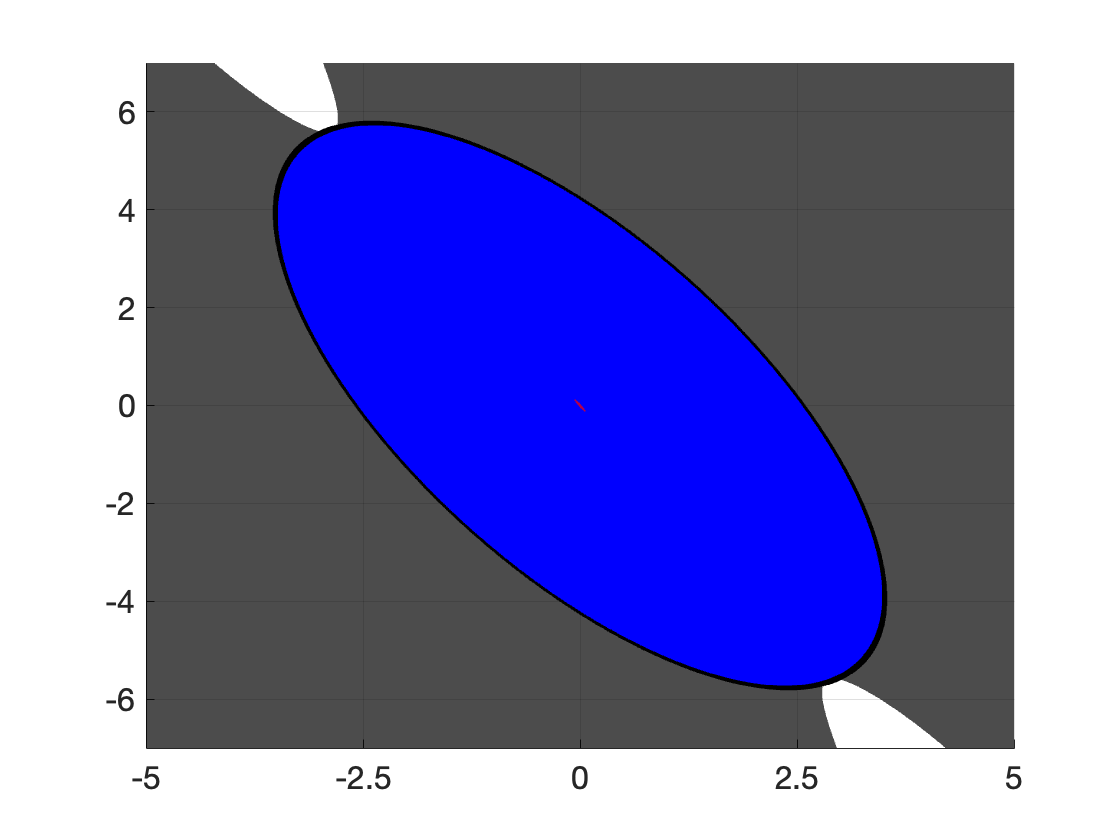}} 
\end{figure}

We solve \eqref{eq:4SDP} 
(recall that \eqref{eq:4SDP} has the same feasible 
set as \eqref{eq:2SDPnoisy}) using the same parameters as in Example 6 but now with 
the average matrices $\underline U_0,\underline Z_0,\underline X_1$, and $\Delta=0.0348$.
We obtain $K = \begin{bmatrix}  -20.9897 & -11.1369  & -9.8222 \end{bmatrix}$. 
Theorem \ref{thm:approx_noise_prob} implies that
$K$ is stabilizing with probability at least $99.48\%$ 
($K$ is indeed stabilizing as $\|\underline D_0\|_2 =  0.0050 < \Delta$). 
The RPI set obtained with $\Delta=0.0348$
is much larger than the one obtained in Example 6 
with the worst-case value $\Delta=\delta \sqrt{T}$;
compare the new Figure \ref{fig:pendulum_RPI_average} with 
Figure \ref{fig:pendulum_RPI}. 
\quad $\blacksquare$ 

\begin{figure}[t]
\normalsize
\caption{Simulation results for the polynomial system of Example 4 with
a disturbance uniformly distributed in $[-0.001,0.001]$ which affects both 
the states. We consider trajectories of length $T=50$ and solve
\eqref{eq:4SDP} with $\lambda_1=\lambda_2=0.1$.
Left: results without averaging. The grey set represents the set $\mathcal X$ in \eqref{eq:RPI_set_X},
while the blue set is the RPI set.
Right: results with averaging ($N=10$). We took $\mu=5\cdot10^{-7}$
which gives $\Delta=0.0052I_2$ and certifies stability with $98.86\%$ probability. 
}
\label{fig:poly_noisy} 
\hspace*{-0.5cm}
{\includegraphics[width=4.8cm]{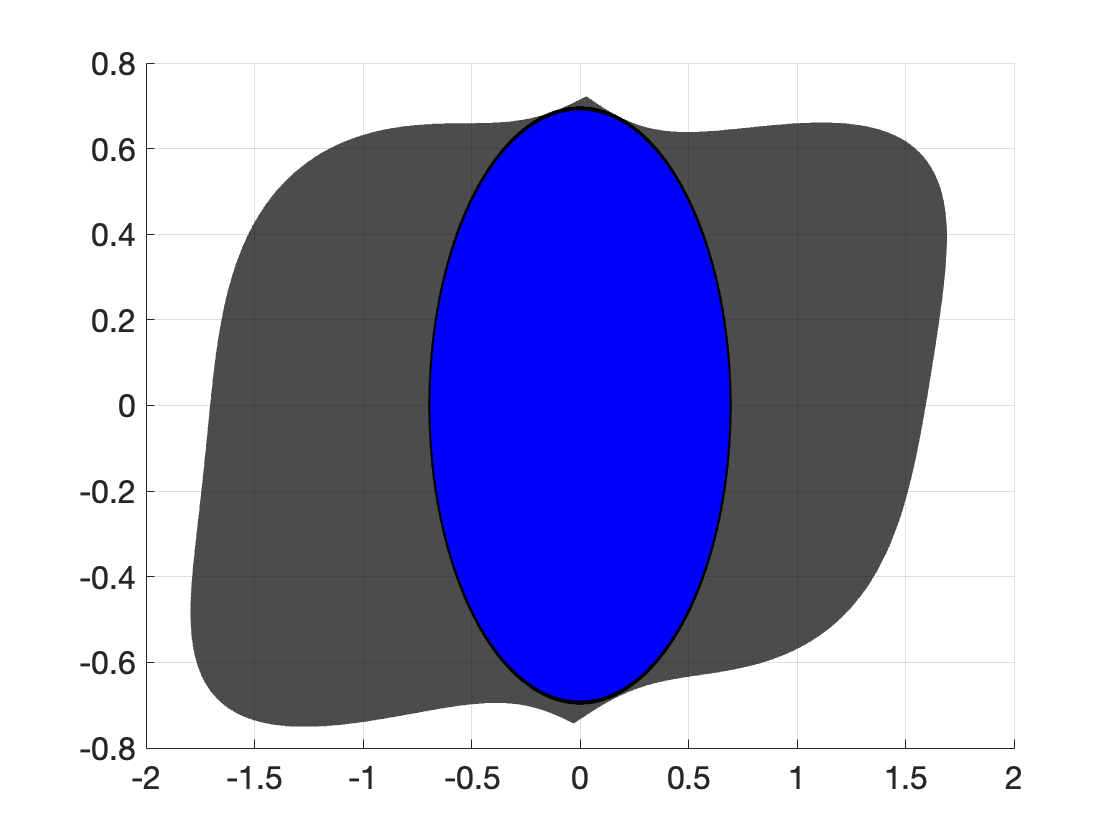}} \hspace*{-0.6cm}
{\includegraphics[width=4.8cm]{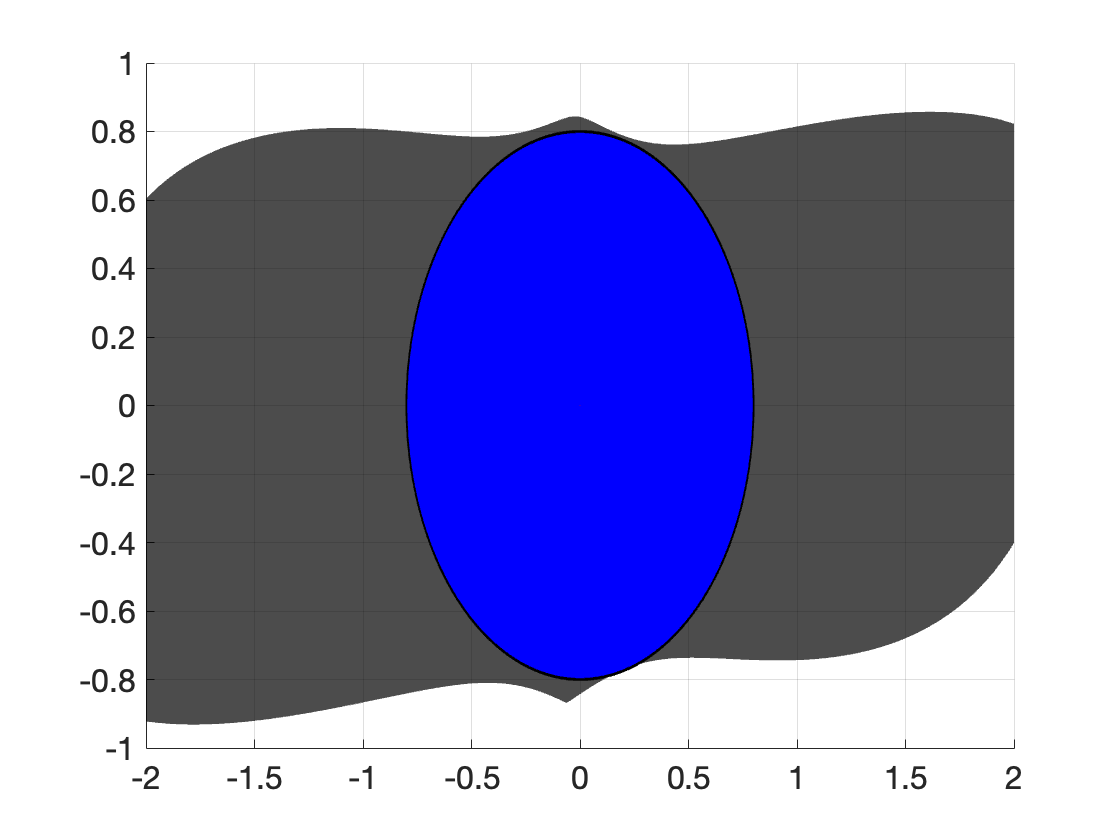}} 
\end{figure}

\textbf{Example 9.} We conclude the section with some simulation results 
for the polynomial system of Example 4. The system has ``more unstable" dynamics
than the pendulum system, and we obtain non-negligible RPI sets 
only for $|d| \leq 0.001$. For the same setting as in Example 4 and 
a disturbance uniformly distributed the SDP \eqref{eq:4SDP} returns the RPI set
in Figure \ref{fig:poly_noisy} (Left). With averaging, we already  
improve the estimate for $N=10$, see Figure \ref{fig:poly_noisy} (Right).
With averaging, we also systematically
obtain non-negligible RPI sets up to $|d| \leq 0.01$. \quad $\blacksquare$

\section{Discussion}\label{sec:disc}


\subsection{Approximate nonlinearity cancellation and ROA size}
Exact nonlinearity cancellation leads to global asymptotic controllers in the case 
no noise is affecting the data used in the design (Theorem \ref{thm:exact}). When an 
exact cancellation of the nonlinearities is not possible, an approximate one should be 
considered, as studied in Theorem \ref{thm:approx}. In general this result returns a 
local asymptotic stabilizer. Here, we would like to stress that this does not 
imply that it does not exist a global stabilizer attaining the same cost as
the feasible solutions of the 
SDP \eqref{eq:2SDP} appearing in Theorem \ref{thm:approx}. We illustrate this point by 
revisiting system \eqref{exmp:5} in Example 4, which was used to demonstrate 
Theorem \ref{thm:approx} and its follow-up, Proposition \ref{prop:RoA}.

We  observe that, were the model of the system known, one could design a global 
asymptotic stabilizer given by
$u = -x_2-0.1 x_1^2-x_1^3 - 0.08 x_1 x_2^2 - 0.016 x_2^4$.
This controller returns a closed-loop system whose linear part $M$ is Schur and 
whose nonlinear part $N$ has norm equal to $0.2$, the optimal value attained by 
the SDP \eqref{eq:2SDP}. Hence, if one would include quartic monomials in $Z(x)$, 
it could be numerically verified whether or not
the global asymptotic stabilizer is a feasible solution 
to the SDP  \eqref{eq:2SDP}. However, there is no analytic guarantee that the SDP 
will return exactly the global stabilizer, and in general it will not. 
This is because the SDP is obtained adopting a quadratic Lyapunov function and
does not currently include a constraint to select a controller that maximizes 
the region of attraction, topics which are left for future research. 

\subsection{Nonlinearity cancellation and coordinate transformations}
 
In model-based design, the possibility of cancelling the nonlinearity is eased by the 
existence of a normal form revealed by a suitable coordinate transformation. In this 
section we comment on how the techniques investigated so far lend themselves to 
be used along with such coordinate transformations obtainable for  systems having 
a uniform relative degree equal to the dimension of the state space.  

Consider the discrete time nonlinear system with output  
\begin{subequations} \label{nonl.dt}
\begin{alignat}{2}
x^+ & = f(x, u)\\
y & =  h(x)
\end{alignat}
\end{subequations}
where $u,y\in \mathbb{R}$ for the sake of simplicity. 
We assume that both the state $x$ and the output $y$ are available for measurements. 
A prior information about the system is that it satisfies
\begin{equation} \label{unif.rel.degree}
\begin{array}{rll}
\displaystyle\frac{\partial h\circ f_0^i\circ f(x,u)}{\partial u} = 0, & \!\!\!\forall (x,u) 
\in \mathbb{R}^{n+1}, &0\le i\le n-2\\[3mm]
\displaystyle\frac{\partial h\circ f_0^{n-1}\circ f(x,u)}{\partial u} \ne 0, & 
\forall (x,u)\in \mathbb{R}^{n+1}&
\end{array} 
\end{equation} 
where $f_0(x)=f(x,0)$, $f_0^d=\underbrace{f_0\circ f_0\circ\ldots \circ f_0}_{\textrm{$d$ times}}$, 
\begin{equation} \label{Phi0}
\begin{bmatrix}
h(x)\\
h\circ f_0(x)\\
\vdots\\
h\circ f_0^{n-1}(x)
\end{bmatrix} =: \Phi_0(x) 
\end{equation} 
is a global coordinate transformation (\cite{monaco1987minimum, isidori1995nonlinear}). 
The transformation $\Phi_0$ depends on the system's dynamics, 
which is not available; nevertheless it can be implemented bearing in mind the interpretation of its 
entries as the value of the output at a given time and at future time instants, namely, at any time $k$, 
we have that 
\begin{equation*}
w(k):=
\begin{bmatrix}
y(k)\\
y(k+1)\\
\vdots\\
y(k+n-1)
\end{bmatrix}=\Phi_0(x(k)) ,
\end{equation*}
so that in the coordinates $w$ the system's dynamics can be written as 
\begin{equation} \label{normal.form}
\!w(k+1)=
\begin{bmatrix}
w_2(k)\\
w_3(k)\\
\vdots\\
w_n(k)
\\
h\circ f_0^{n-1}\circ f(x(k), u(k))
\end{bmatrix}\!,
\,y(k)  =  w_1(k)
\end{equation}

Note that the last entry of the vector field on the right-hand side has been deliberately left to 
depend on the original state $x$ rather on the new one $z$, which turns out to be useful to obtain 
a {\em causal} control policy. 
The point of this transformation is that, were the system's dynamics known, one could design 
a static feedback controller that stabilizes the system via exact  nonlinearity cancellation. 
When the dynamics are unknown, one can still achieve exact nonlinearity cancellation 
by modifying the techniques proposed in Section \ref{subsec:exact}, provided that the 
following assumption holds:

\begin{assumption}\label{asspt:regressor.rel.degree.n}
A vector-valued  function 
$Q: \mathbb R^n \rightarrow \mathbb R^{S-n}$
is known for which $h\circ f_0^{n-1}\circ f(x, u) =  a^\top Q(x)+bu$ for some (unknown) quantities 
$a\in  \mathbb{R}^{S}$, $b\in \mathbb{R}\setminus \{0\}$. 
\end{assumption} 

Asking for $h\circ f_0^{n-1}\circ f(x, u)$ to take this specific form is clearly demanding, 
but one can in principle collect the discrepancy between $h\circ f_0^{n-1}\circ f(x, u)$ and 
$a^\top Q(x)+bu$ into a mismatch function and treat it as a disturbance,
analogously to what has been discussed in Section \ref{subsec:neglected}.

Under the assumption above, a controller can be designed following the construction in the previous 
subsection with suitable modifications. We start defining the matrix of input samples $U_0$ 
as in \eqref{eq:data1}, and 
\begin{subequations}
\begin{alignat}{4}
& W_0 := \begin{bmatrix} w(0) & w(1) & \cdots & w(T-1)  \end{bmatrix} 
\in \mathbb R^{n \times T} \,, \nonumber 
\\
& W_1 := \begin{bmatrix} w(1) & w(2) & \cdots & w(T)  \end{bmatrix} 
\in \mathbb R^{n \times T} \,, \nonumber \\
& Q_0 := \begin{bmatrix} Q(x(0))& Q(x(1)) & \cdots & Q(x(T-1))  
\end{bmatrix}\in \mathbb R^{(S-n) \times T}, \nonumber\\
& Z_0 := \begin{bmatrix} W_0^\top & Q_0^\top
\end{bmatrix}^\top \in \mathbb R^{S \times T},
\end{alignat}
\end{subequations}
which satisfy the identity $W_1= A_c W_0 +B_c(a^\top Q_0 + b U_0)$, where the pair $(A_c, B_c)$ is in the 
Brunovsky canonical form \cite{brunovsky1970classification}. 
Note that since both the state $x$ and the output $y$ are assumed to 
be available for measurements, the matrices of data $W_0, W_1, Q_0$ are known. 
In particular, the matrix $W_0$ (similarly for $W_1$) comprises output samples:
\begin{eqnarray*}
W_0 = \begin{bmatrix}
y(0) & y(1) & \ldots & y(T-1)\\
y(1) & y(2) & \ldots & y(T)\\
\vdots & \vdots & \ddots & \vdots\\
y(n-1) & y(n) & \ldots & y(n+T-2)\\
\end{bmatrix} .
\end{eqnarray*}
We have the following result. 

\begin{corollary}\label{cor:feedb_lin}
Consider the nonlinear system with output \eqref{nonl.dt}. 
Assume that conditions \eqref{unif.rel.degree} hold and that the map $\Phi_0$ in \eqref{Phi0} 
is a global coordinate transformation. If there exist decision variables $G_1 \in \mathbb R^{T \times n}$,
$k_1 \in \mathbb R$, and $G_2 \in \mathbb R^{T \times (S-n)}$ such that
\begin{subequations}
\label{eq:SDP-fl}
\begin{alignat}{2}
& Z_0 G_1 = \begin{bmatrix} I_n \\ 0_{(S-n) \times n} \end{bmatrix} \,,
\label{eq:SDP1-fl} \\
& W_1 G_1 = A_c + B_c \, \big[ k_1 \,\,\,\, \underbrace{0 \,\,\,\, \cdots \,\,\,\, 0}_{n-1 \text{ times}} \big] \,, 
\label{eq:SDP2-fl} \\[-0.2cm]
& k_1 \in (-1,1) \,, \\
& Z_0 G_2 = \begin{bmatrix} 0_{n \times (S-n)} \\ I_{S-n} \end{bmatrix} \,,
\label{eq:SDP4-fl} \\
& W_1 G_2 = 0_{n \times (S-n)} \,, \label{eq:SDP5-fl} 
\end{alignat}
\end{subequations} 
then $u=K\left[\begin{smallmatrix}
w\\ Q(x)
\end{smallmatrix}\right]$, with $K=U_0 G$,  linearizes the closed-loop system 
and renders the origin a globally asymptotically stable equilibrium. \quad $\Box$
\end{corollary}

{\it Proof.} 
Conditions \eqref{eq:SDP1-fl}, \eqref{eq:SDP4-fl} along with the definition of the controller 
gain $K$, show that the identity \eqref{eq:GK} holds. Thus, the closed-loop system is of 
the form 
\begin{subequations} 
\begin{alignat}{2} 
w^+ &= A_c w + B_c (a^\top Q(x) + bu) \\
&= A_c w + B_c (a^\top Q(x) + bU_0G \left[\begin{smallmatrix}
w\\ Q(x) \end{smallmatrix}\right]) \\
&= W_1G \left[\begin{smallmatrix} w\\ Q(x) \end{smallmatrix}\right]  = W_1G_1w
\end{alignat}
\end{subequations}
where the third equality follows from the identities
$B_c b U_0 G = W_1 G  - A_c W_0 G - B_c a^\top Q_0 G$,
\eqref{eq:SDP1-fl} and \eqref{eq:SDP4-fl}, and the last one from 
\eqref{eq:SDP5-fl}.
Hence, the controller $u=K   \left[\begin{smallmatrix} w\\ Q(x)
\end{smallmatrix}\right]$ linearizes the closed-loop system. Finally, by \eqref{eq:SDP2-fl}, 
the closed-loop system coincides with  $w^+=(A_c + B_c \begin{bmatrix} k_1 & 0 & \cdots & 
0\end{bmatrix})w$, where the matrix $A_c + B_c \begin{bmatrix} k_1 & 0 & \cdots & 0\end{bmatrix}$ 
is Schur since  all its eigenvalues are given by the solutions of the equation $\lambda^n=(-1)^n k_1$ 
and $|k_1|<1$.  \quad $\blacksquare$

The control law only uses the variables $y, x$  
and as such it is implementable. In fact, bearing in mind \eqref{eq:SDP1-fl} and \eqref{eq:SDP4-fl},
the identity $W_1G= A_c W_0G +B_c(a^\top Q_0G + b U_0G)$ is equivalent to 
\begin{equation*}
\begin{array}{l}
\begin{bmatrix}
A_c +B_c \begin{bmatrix} k_1 & 0 & \cdots & 0\end{bmatrix} & 0_{n\times (S-n)}
\end{bmatrix}= 
\begin{bmatrix} A_c & 0_{n\times (S-n)}\end{bmatrix}\\[0.1cm]
+ 
B_c \begin{bmatrix} 0_{n\times n}  & a^\top \end{bmatrix} +
B_c \,b \,U_0G
\end{array}
\end{equation*}
from which we deduce that $U_0G = b^{-1} 
\left[\begin{smallmatrix} \left[\begin{smallmatrix} k_1 & 0 & \cdots & 0\end{smallmatrix}\right] 
& -a^\top \end{smallmatrix}\right]$, that is $U_0 G_1w$ only depends on the first component of 
$w$, which is the output $y$.

\textbf{Example 10.} Consider the polynomial system
\begin{subequations} 
\label{eq:sys_example1a}
\begin{alignat}{2}
& x_1^+ =  x_2^2 + x_1^3 + u \label{eq:sys_example1a1} \\   
& x_2^+ = 0.5 x_1 + 0.2 x_2^2\\
& y = x_2
\end{alignat}
\end{subequations}
Exact cancellation based on Theorem \ref{thm:exact} is not possible for this system. 
On the other hand, the conditions of Corollary \ref{cor:feedb_lin} hold. 

In particular, notice that  
\begin{equation*}
h\circ f_0^{n-1}\circ f(x, u) =  \frac{1}{20} x_1^2+ \frac{1}{2} x_2^2+\frac{1}{2} x_1^3 + 
\frac{1}{25} x_1 x_2^2+ \frac{1}{125} x_2^4+ \frac{1}{2}u.
\end{equation*}
Hence, if we choose
\begin{equation*}
\begin{array}{l}
Q(x) = \left[
x_1^2 \; \;  x_2^2  \; \;  x_1 x_2  \; \;  x_1^3  \; \;  x_2^3  \; \;  x_1x_2^2  \; \;  x_1^2 x_2  \; \;  x_1^4 
\; \; x_2^4  \; \;  x_1x_2^3  \; \; \right.
\\
\hspace{4cm} \left.x_1^2 x_2^2   \; \;   x_1^3 x_2 \right]
\end{array}
\end{equation*}
then Assumption \ref{asspt:regressor.rel.degree.n} is satisfied. The choice of 
such a $Q(x)$ can be guided by some prior knowledge, namely that the nonlinearity 
in the last equation of the system in the new coordinates is a polynomial of degree no 
larger than $4$.  On the other hand, the exclusion of $x$ from $Q(x)$ 
is suggested by the fact that, if this were
not the case, then the matrix $Z_0$ would be rank deficient 
(this is a test that can be carried out from the collected data). This is because each 
column $i$ of $W_0$ is equal to $[y(i-1)\;\; y(i)]^\top=[x_2(i-1)\;\; 0.5 x_1(i-1) + 0.2 x_2(i-1)^2]^\top$ 
and it would be  expressible as a linear combination of the entries 
of column $i$ of $Q_0$ if the latter would include $x$. 

Applying Corollary \ref{cor:feedb_lin}, we find that the SDP \eqref{eq:SDP-fl} is feasible and returns 
the solution $k_1=0.372$ and 
\begin{equation*}
\begin{array}{l}
K=\left[
0.7423  \; \;  0  \; \;  -0.1  \; \;   -1  \; \;   0  \; \;  -1  \; \;   0  \; \;   -0.08   \; \;   0    \; \;      0  \right. \\
\hspace{4cm} \left. -0.016          \; \; 0  \; \;    0  \; \;  0 \right]
\end{array}
\end{equation*}
which linearizes the closed-loop system in the coordinates $w$,
and renders the origin a globally asymptotically stable equilibrium.  \quad $\blacksquare$

\section{Conclusions}
 
We have introduced a method to design Lyapunov-based stabilizing controllers 
for nonlinear systems from data, which reduces the design to the solution of data-dependent 
SDP.  The method is certified to provide a solution in the presence 
of perturbed data as well as estimates of the region of attraction of the closed-loop system. 
Both deterministic and stochastic perturbations on the data are studied. We also extended 
the results to deal with the presence of neglected nonlinearities. 
Possible future research should focus on output feedback control design, the inclusion 
of criteria to maximize the region of attraction and the design of more general 
(non quadratic) Lyapunov functions. 

\appendix 
  
\subsection{A parametrization of all stabilizing and linearizing feedback controllers}
\label{subsec:param:exact}
 
Suppose that $[\begin{smallmatrix} U_0 \\ Z_0 \end{smallmatrix}]$ has full row rank. In this case,
we can prove that any stabilising and linearising feedback controller can be 
parametrised as in \eqref{eq:K_SDP} for some $Y_1,P_1,G_2$
satisfying \eqref{eq:SDP}. Note in particular that this implies that the SDP is feasible. 
This result is as a generalization of \cite[Theorem 3]{de2019formulas}
where an analogous result for linear system is provided under the
condition that $[\begin{smallmatrix} U_0 \\ X_0 \end{smallmatrix}]$ 
has full row rank. In the linear case, the latter condition reduces to 
a design condition for controllable dynamics, see \cite[Theorem 1]{fundamental-lemma},
\cite[Theorem 1]{henk-fl-multiple-traj}.
To the best of our knowledge, no analogous design conditions exists 
for nonlinear systems.

\emph{Proof of Theorem \ref{thm:param:exact}}. Consider any 
stabilizing and linearizing feedback controller $K$. We have
{\setlength\arraycolsep{2.0pt}
\begin{eqnarray} \label{eq:paramK}
A+BK 
= X_1 G
\end{eqnarray}}%
for some $G \in \mathbb R^{T \times S}$ satisfying \eqref{eq:GK}. Note that
$G$ exists as 
$[\begin{smallmatrix} U_0 \\ Z_0 \end{smallmatrix}]$ has full row rank
by hypothesis. By partitioning $K=[\begin{matrix} \overline K & 
\hat K \end{matrix}]$ with $\overline K \in \mathbb R^{m \times n}$
and $G=[\begin{matrix} G_1 & G_2 \end{matrix}]$ with 
$G_1 \in \mathbb R^{T \times n}$, we have $X_1 G_1 = \overline A + B \overline K$
and $X_1 G_2 = \hat A + B \hat K = 0$,
where the matrix $X_1 G_1$ is Schur 
and $X_1 G_2 = 0$ by the assumption that $K$ is stabilizing and linearizing.
Hence, there exists a matrix $P_1 \succ 0$ such that 
$(X_1 G_1)^\top P_1^{-1} X_1 G_1 - P_1^{-1} \prec 0$.
This implies $(X_1 Y_1)^\top P_1^{-1} X_1 Y_1 - P_1 \prec 0$ with $Y_1 = G_1 P_1$,
which is the stability constraint in \eqref{eq:SDP2}. Since $Z_0G=I_S$ and $Y_1 = G_1 P_1$ we have
\begin{equation} \label{eq:app1}
Z_0 \begin{bmatrix} Y_1 & G_2 \end{bmatrix}  = 
\begin{bmatrix} P_1 & 0_{n \times (S-n)}  \\ 0_{(S-n) \times n} & I_{S-n} 
\end{bmatrix},
\end{equation}
which matches the constraints \eqref{eq:SDP1} and \eqref{eq:SDP4}.
Thus, all the constraints in \eqref{eq:SDP} are satisfied,
hence the program is feasible.

As for the form of the controller, by \eqref{eq:GK} we have $K=U_0G$
which in terms of $Y_1,G_2$ reads as \eqref{eq:K_SDP}. \quad $\blacksquare$

\subsection{A parametrisation of all (locally) stabilising feedback controllers}
\label{subsec:param:approx}

\emph{Proof of Theorem \ref{thm:param:approx}.} 
The identity \eqref{eq:paramK} is still valid because independent 
of the properties of $K$. Furthermore, we can still write $X_1 G_1 = \overline A + B \overline K$
and $X_1 G_2 = \hat A + B \hat K$.
(The only difference with respect to Theorem \ref{thm:param:exact} is that now
$X_1 G_2$ might be different from zero.)
Observe now that, by assumption, $X_1 G_1$ is Schur. 
Hence, there exists a matrix $P_1 \succ 0$ 
such that $(X_1 G_1)^\top P_1^{-1} X_1 G_1 - P_1^{-1} \prec 0$.
By defining $Y_1 = G_1 P_1$, this is equivalent to \eqref{eq:2SDP2}.
Finally, recalling that $Z_0G=I_S$, we have again the 
identity \eqref{eq:app1}.
Thus, all the constraints in \eqref{eq:2SDP} are satisfied
and the program is feasible.

As for the form of the controller, by \eqref{eq:GK} we have $K=U_0G$
which in terms of $Y_1,G_2$ reads as \eqref{eq:K_SDP}. \quad $\blacksquare$

\subsection{Proof of Lemma \ref{lem:Petersen}} \label{sec:petersen}
 
Lemma \ref{lem:Petersen} is a direct consequence of the following result. 

\begin{lem} \label{lem:Petersen_aux}
Let $B \in \mathbb R^{n \times p}$, $C \in \mathbb R^{q \times n}$ be given matrices, and 
let $\mathcal D:=\{D \in \mathbb R^{q \times p}: D D^\top \preceq \Delta \Delta^\top\}$. Then, 
for arbitrary $\epsilon > 0$ it holds that 
\begin{equation*}
BD^\top C + C^\top D B^\top \preceq 
\epsilon^{-1} B B^\top + \epsilon C^\top \Delta \Delta^\top C
\quad \forall D \in \mathcal D
\end{equation*} 
\end{lem} 

\emph{Proof}. A completion of squares
\begin{equation*}
\left( \sqrt{\epsilon^{-1}} B -
\sqrt{\epsilon} C^\top D \right) 
\left( \sqrt{\epsilon^{-1}} B - 
\sqrt{\epsilon} C^\top D \right)^\top
\succeq 0
\end{equation*} 
gives the result. \quad $\blacksquare$

\emph{Proof of Lemma \ref{lem:Petersen}.}
Let \eqref{eq:Petersen} hold. By a Schur complement, this is equivalent to 
\begin{eqnarray*} \label{eq:app13} 
\left[ \begin{array}{cc}
P_1-\Omega & (X_1Y_1)^\top \\ X_1 Y_1 & P_1
\end{array} \right] - 
\epsilon^{-1} \underbrace{\left[
\begin{array}{c}
Y_1^\top \\ 
0_{n \times T} 
\end{array}
\right]}_{:=B} \left[
\begin{array}{cc}
Y_1 &  
0_{T \times n} 
\end{array}
\right] \nonumber \\ - \epsilon \underbrace{\left[
\begin{array}{c}
0_{n \times s}  \\ 
E
\end{array}
\right]}_{:=C^\top} \Delta \Delta^\top \left[
\begin{array}{cc}
0_{s \times n} &  
E^\top 
\end{array}
\right]  \succ 0 
\end{eqnarray*} 
An application of Lemma \ref{lem:Petersen_aux} gives
\begin{eqnarray*} \label{eq:} 
\left[ \begin{array}{cc}
P_1-\Omega & (X_1Y_1)^\top \\ X_1 Y_1 & P_1
\end{array} \right] -
\left[
\begin{array}{c}
Y_1^\top \\ 
0_{n \times T} 
\end{array}
\right] D^\top \left[
\begin{array}{cc}
0_{s \times n} &  
E^\top \end{array}
\right] \nonumber \\ - \left[
\begin{array}{c}
0_{n \times s}  \\ 
E \end{array}
\right] D \left[
\begin{array}{cc}
Y_1 &  
0_{T \times n}
\end{array}
\right] \succ 0 \nonumber \\
\quad \forall D \in \mathcal D 
\end{eqnarray*} 
or, equivalently,
\begin{eqnarray} \label{eq:app15} 
\left[
\begin{array}{cc}
P_1-\Omega & Y_1^\top (X_1-ED)^\top \\
(X_1-ED) Y_1 & P_1
\end{array}
\right] \succ 0 \quad \forall D \in \mathcal D
\end{eqnarray} 

This is equivalent to \eqref{eq:2SDP_noisy} after another Schur complement,
and this gives the result. \quad $\blacksquare$

\bibliographystyle{IEEEtran}
\bibliography{refs}

\end{document}